\documentclass[twocolumn,preprint]{aastex62}
\def\gsim{\ \raise 3pt \hbox{$\rangle$} \kern -8.5pt \raise -2pt \hbox{$\sim$}\ }
\usepackage{graphicx,natbib,color,amsmath,subfigure}
\usepackage{ulem}
\newcommand{\blank}[1]{}
\usepackage{color}

\newcommand{\vect}[1]{\mbox{\boldmath $#1$}}   

\begin{document}
\title{Gyroresonance and free-free radio emissions from multi-thermal multi-component plasma}

\author[0000-0001-5557-2100]{Gregory D. Fleishman}
\affil{Center For Solar-Terrestrial Research, New Jersey Institute of Technology, Newark, NJ 07102\\
		Ioffe Institute, Polytekhnicheskaya, 26, St. Petersburg, 194021, Russia}
\author[0000-0001-8644-8372]{Alexey A. Kuznetsov}
\affil{Institute of Solar-Terrestrial Physics, Irkutsk, 664033, Russia}
\author[0000-0002-9325-9884]{Enrico Landi}
\affil{Department of Climate and Space Sciences and Engineering, University of Michigan, USA}

\begin{abstract}

Thermal plasma of solar atmosphere includes a wide range of  temperatures. This plasma is often quantified, both in observations and models, by a differential emission measure (DEM). DEM is a distribution of the thermal electron  \textit{density square} over temperature. In observations, the DEM is computed along a line of sight, while in the modeling---over an elementary volume element (voxel). This description of the multi-thermal plasma is convenient and widely used in the analysis and modeling of extreme ultraviolet emission (EUV), which has an optically thin character. However, there is no corresponding treatment in the radio domain, where optical depth of emission can be large, more than one emission mechanism are involved, and plasma effects are important. Here, we extend the theory of the thermal gyroresonance and free-free radio emissions in the classical mono-temperature Maxwellian plasma to the case of a multi-temperature plasma. The free-free component is computed using the DEM and temperature-dependent ionization states of coronal ions, contributions from collisions of electrons with neutral atoms, exact Gaunt factor, and the magnetic field effect. For the gyroresonant component, another measure of the multi-temperature plasma is used 
which describes the distribution of the thermal electron \textit{density} over temperature. We give representative examples demonstrating important changes in the emission intensity and polarization due to considered effects.  The theory is implemented in available computer code.

\end{abstract}


\section{Introduction}

Quiescent, slowly varying radio emission from the Sun is formed by two distinct emission processes: free-free emission (bremsstrahlung) due to collisions of thermal electrons with plasma ions or atoms and gyro emission due to thermal electron gyration in the ambient magnetic field. Theoretically, these mechanisms are well understood at both classical and quantum levels: emission from a plasma with given electron density  $n_e$, temperature $T$, chemical composition, and magnetic field $\mathbf{B}$ can be computed precisely.

However, application of the theory to numerical modeling of a realistic, nonuniform  solar or stellar atmosphere requires that the volume resolution elements (voxels) represent uniform volume(s) with single, well-defined values of $n_e$, $T$, and $\mathbf{B}$. In principle, this can be achieved by devising very-high-resolution models with a very small voxel size, which is the case of some full-fledged MHD models \citep{2011A&A...531A.154G,2016A&A...585A...4C}. Such generic models are limited to a relatively small region of the atmosphere and cannot encompass the entire active region or a combination of them.

In contrast, data-constrained or data-driven models have typically much lower spatial resolution; in particular, because the input data used to build such models are lower resolution. In such cases, each individual voxel represents a nonuniform volume that contains components with different temperatures and densities distributed over the voxel. Having these distributions we can compute their moments to derive averaged values---mean density, mean squared density, and mean temperature in the voxel. However, in a general case, emission computed using those averaged values will differ from emission computed from the truly nonuniform volume. This is broadly recognized in the analysis and modeling of the optically thin extreme ultraviolet (EUV) and soft X-ray (SXR) data sensitive to the square of the electron number density---emission measure (EM), where  explicit account of the plasma nonuniformity is crucial. This nonuniformity is typically accounted in the form of the differential emission measure (DEM)---a distribution of the plasma over the temperature (see below for the quantitative view).

No similar framework is available to compute radio emission, although it is demanded for analysis and modeling the currently available data and needed to fairly combine  SXR, EUV, and radio models. Extending the theory of the free-free and gyro emission from a uniform to a nonuniform voxel in the radio domain is challenging because they involve both optically thin and thick regimes unlike EUV/SXR emissions, which are optically thin. In addition, while the free-free emission (mainly) depends on the EM (square of the density)---same as EUV and SXR, the gyro emission depends on the density itself. Thus, an additional measure, which we call the differential 
density metrics (DDM), is needed to compute it. Free-free emission also requires some additional information about the plasma ionization because collisions between electrons and ions depend on the ion charge. Thus, precise computation of the free-free emission from a solar or stellar atmosphere requires accurate information about the heavy ions and similarly accurate treatment of other factors affecting the free-free emission---the Gaunt factor (Coulomb logarithm) and the factor accounting for the ambient magnetic field effect. In this paper we develop the theory that accounts for all these mentioned effects in a multi-thermal plasma of a  solar or stellar atmosphere and implement this theory in the computer codes that solve for the radiation transfer numerically.



\section{Radiation transfer in a multi-temperature plasma}
\label{S_transfer}


The equation of the radiation transfer without scattering has the form: 
\begin{equation}
\label{Eq_transfer}
 \frac{d{\cal J}^{\sigma}_{ f}(\vect{r},t)}{v_g^{\sigma}dt}
 =j^{\sigma}_{f}(\vect{r},t) -
 \varkappa^{\sigma}(\vect{r},t){\cal J}^{\sigma}_{f}(\vect{r},t),
\end{equation}
where ${\cal J}^{\sigma}_{f}(\vect{r},t)$ [erg\,cm$^{-2}$\,s$^{-1}$\,Hz$^{-1}$\,Sr$^{-1}$] 
is the radiation intensity of either ordinary (O; $\sigma=1$) or extraordinary (X; $\sigma=-1$) wave mode for a given radiation direction, $j^{\sigma}_{f}(\vect{r},t)$ and $\varkappa^{\sigma}(\vect{r},t)$ are the corresponding emissivity and absorption coefficient for this direction,
$\frac{d}{dt}=\frac{\partial}{\partial t} + \vect{v}_g^{\sigma}\frac{\partial}{\partial \vect{r}}$ is the full derivative over time, $\vect{v}_g^{\sigma}=\partial \omega / \partial \vect{k}$ is the group velocity different for the eigen-modes \citep{Fl_etal_2002}.
A numerical solution of this equation relies on representation of the line of sight (LOS) as a sequence of (quasi-)uniform volume elements (voxels). The solution of the radiation transfer equation through a single voxel is

\begin{equation}\label{Eq_Inten_voxel}
{\cal J}_{f,n}^\sigma={\cal J}_{f,n-1}^\sigma\exp(-\tau_{n}^\sigma)+\frac{j_{f,n}^\sigma}{\varkappa_{n}^\sigma}\left(1-\exp(-\tau_{n}^\sigma)\right),$$$$ \ \ {\rm [erg\ cm^{-2} s^{-1} Hz^{-1} Sr^{-1}]} ,
\end{equation}
where ${\cal J}_{f,n-1}^\sigma$ and ${\cal J}_{f,n}^\sigma$ are the emission intensities entering and exiting  voxel $\#n$,  $\tau^{\sigma}_n=\varkappa^{\sigma}_n\Delta r$ is the optical depth of the voxel at a given frequency, and $\Delta r$ is the voxel length along the LOS. The final result of the transfer of the radiation through a LOS composed of $N$ voxels, ${\cal J}_{f}^\sigma$, is obtained with sequential use of Eq.\,(\ref{Eq_Inten_voxel}) $N$ times starting from $n=1$ and ending at $n=N$.

For practical applications we convert the radiation intensity ${\cal J}_{f}^\sigma$ to two other widely used physical measures---the brightness temperature $T_B^\sigma$ of the $\sigma$-mode emission 
\begin{equation}
\label{Eq_T_br2J}
 {\cal J}_{f}^\sigma= 2 \frac{f^2}{ c^2 }k_B T_B^\sigma\ \ {\rm [erg\ cm^{-2} s^{-1} Hz^{-1} Sr^{-1}]} ,
\end{equation}
such as
\begin{equation}
\label{Eq_T_br_def}
 T_B^\sigma=
  \frac{ c^2 }{2k_B f^2}
  {\cal J}_{f}^\sigma, \ \ {\rm [K]}
\end{equation}
and the radio flux
from a source at the Sun located at $\mathcal{R}$=1\,AU=$1.496\times 10^{13}$\,cm from the observer expressed in solar flux units (sfu; 1\,sfu = 10$^4$\,Jy= 10$^{-19}$\,erg~s$^{-1}$\,cm$^{-2}$\,Hz$^{-1}$)
\begin{equation}
S_f^\sigma= 10^{19}{A\over \mathcal{R}^2}{\cal J}_{f}^\sigma\quad {\rm [sfu]},
\end{equation}
\noindent where
$A/\mathcal{R}^2$ is the solid angle
subtended by a source with visible area $A$. The Stokes $I$ measures represent the sums of those for the two modes: ${\cal J}_{f} = {\cal J}_{f}^X + {\cal J}_{f}^O$ and
$S_{f} = S_{f}^X + S_{f}^O$.  Accordingly, {although temperature is an} intensive {property} 
\citep[see, e.g.,][]{FT_2013}, the brightness temperature is defined {here} as an extensive (additive) {property} $T_B = T_B^X + T_B^O$. In particular, for the unpolarized optically thick emission from a Maxwellian plasma with the temperature $T$ we have $T_B=T$ and $T_B^X = T_B^O = T/2$.

If the numerical grid is reasonably fine, then each voxel can be considered as a truly uniform, single-temperature, volume  so that the GR and free-free emissivity and absorption coefficient computed for a uniform single-temperature Maxwellian plasma apply. In practice, however, the voxel size might be much larger than the typical size of a single-temperature volume. Thus, we are to compute the radio emission from a voxel, which might contain multi-temperature plasma, even though it could be treated as a uniform volume from all (most of the) other perspectives. We assume that each voxel can be considered as a ``statistically uniform'' volume, such as the thermal properties of each macroscopic fraction of the voxel are the same as those of the entire voxel. In this case we can safely use the uniform-source solution

\begin{equation}\label{Eq_Inten_voxel_mean}
{\cal J}_{f,n}^\sigma={\cal J}_{f,n-1}^\sigma\exp(-\left<\tau_{n}^\sigma\right>)+\frac{\left<j_{f,n}^\sigma\right>}{\left<\varkappa_{n}^\sigma\right>}\left(1-\exp(-\left<\tau_{n}^\sigma\right>)\right),
\end{equation}
of radiative transfer equation (\ref{Eq_transfer}), where $\left<j^{\sigma}_{f,n}\right>$ and $\left<\varkappa^{\sigma}_{n}\right>$ are the emissivity and the absorption coefficients averaged over the voxel volume,  and
$\left<\tau^{\sigma}_{n}\right> = \left<\varkappa^{\sigma}_{n}\right>\Delta r$ is the  volume-averaged optical depth of the voxel.

\section{Description of multi-temperature plasma} 

Various ingredients of the emissivities and the absorption coefficients, described in detail below, can depend either on $n_e$ or $n_e^2$, so we need to define distributions of those two quantities over the voxel volume. We start from a frequently used value $n_e^2$, which is directly linked to a well-known plasma emission measure. To make our quantities independent of the voxel volume $V$, we adopt a volume-normalized definition for the EM{, which, in this case, is equivalent to the volume-averaged square of the electron density $\left<n_e^2\right>$}:

\begin{equation}
\label{Eq_EM_over_V}
 EM=\left<n_e^2\right>=\frac{1}{V} \int n_e^2(\mathbf{r}) dV, \quad {\rm [ cm^{-6}]}.
\end{equation}
It is customary to replace the integral over the voxel volume by the integration over the range of temperatures available in the volume such as

\begin{equation}
\label{Eq_EM_over_dem}
 EM=\frac{1}{V} \int n_e^2(T) \frac{dV}{dT} dT = \int \xi(T) dT,  \quad {\rm [ cm^{-6}]},
\end{equation}
where
\begin{equation}
\label{Eq_dem_def}
 \xi(T)=  \frac{n_e^2(T)dV}{VdT},  \quad {\rm [ cm^{-6}\, K^{-1}]}
\end{equation}
is the differential emission measure (DEM). {Note that this definition differs from the standard DEM definition by the volume normalization.} The average square of the electron density is related to the DEM by
\begin{equation}
\left<n_e^2\right>=\int\xi(T)\,\mathrm{d}T,
\end{equation}
and the DEM-averaged temperature $\left<T_2\right>$ is given by
\begin{equation}
\left<T_2\right>=\frac{\int T\xi(T)\,\mathrm{d}T}{\left<n_e^2\right>}.
\end{equation}

Similarly, we define the {mean density} 

\begin{equation}
\label{Eq_DM_over_V}
 \left<n_e\right>=\frac{1}{V} \int n_e(\mathbf{r}) dV, \quad {\rm [ cm^{-3}]},
\end{equation}
of the thermal electrons in the voxel.
Using again the integration over the range of temperatures, we obtain

\begin{equation}
\label{Eq_DM_over_ddm}
 \left<n_e\right>=
 \frac{1}{V} \int n_e(T) \frac{dV}{dT} dT = \int \nu(T) dT,  \quad {\rm [ cm^{-3}]},
\end{equation}
where
\begin{equation}
\label{Eq_ddm_def}
 \nu(T)=  \frac{n_e(T)dV}{VdT},  \quad {\rm [ cm^{-3}\, K^{-1}]}
\end{equation}
is
{a new metrics that describes distribution of the electron number density over the temperature in the given voxel, which we will call}
the differential {density metrics} 
(DDM) {for short}. Hence the DDM-averaged temperature $\left<T_1\right>$ is given by
\begin{equation}
\left<T_1\right>=\frac{\int T\nu(T)\,\mathrm{d}T}{\left<n_e\right>}.
\end{equation}
 We note that the DDM- and DEM-averaged electron densities and temperatures are different from each other  in a general case.

\begin{figure*}
\resizebox{8.5cm}{!}{\includegraphics{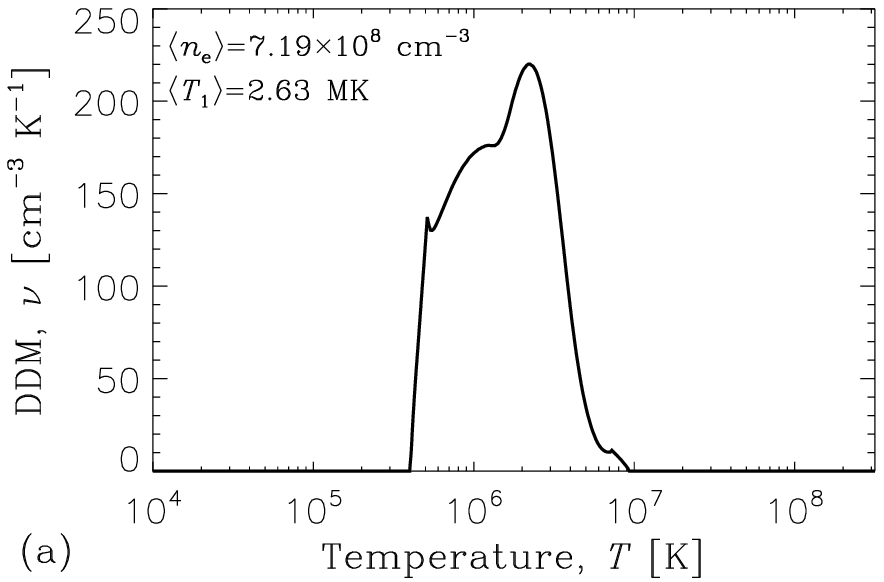}}\resizebox{8.5cm}{!}{\includegraphics{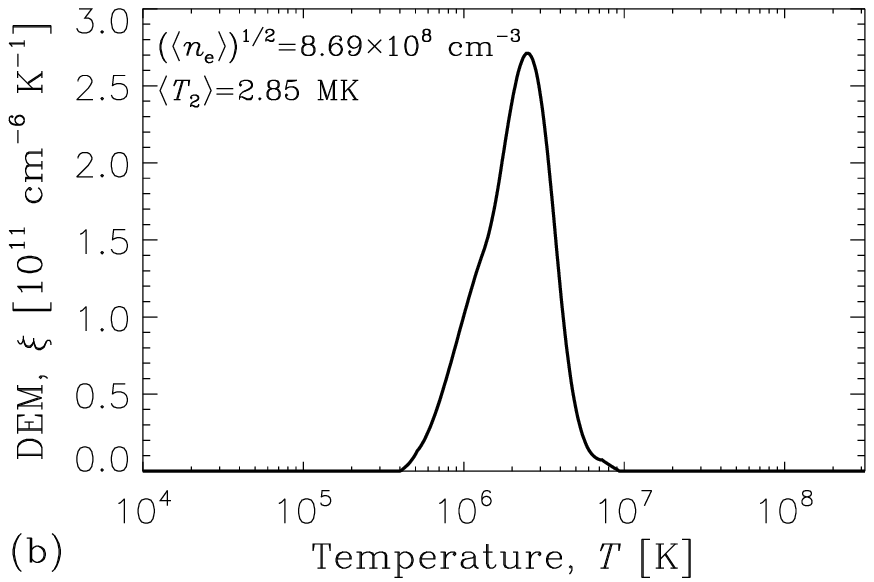}}
\caption{ Model DDM and DEM distributions (courtesy of Jim Klimchuk) used in the simulations.}\label{DEMDDMexample}
\end{figure*}

Figure \ref{DEMDDMexample} shows an example of DDM and DEM distributions obtained from an updated EBTEL code  \citep[courtesy of J. Klimchuk; private communication; for the original EBTEL code see][and references therein]{2008ApJ...682.1351K,2012ApJ...752..161C}; these distributions will be used below (in Sections \ref{FFfromDEM} and \ref{GRfromDDM}) to compute the model emission spectra. The corresponding distribution moments (averaged values of the density and temperature for the DDM and DEM distributions, respectively) 
are printed in the panels; the values inferred from DEM are slightly higher than those inferred from DDM.

\section{Free-free emission from Maxwellian distributions}

Free-free emission from the ionized Maxwellian plasma is described by the following equations \citep[e.g., Section 10.1.1 in][]{FT_2013} for the emissivity\footnote{Note a typo---$(2\pi)^{3/2}$ instead of $(2\pi)^{1/2}$ in the denominator of Eqn.\,(10.4) of \citet{FT_2013}.}

\begin{equation}\label{Eq_emis_ff_Max_a}
j_{f,ff}^{M,\sigma}=\sum_i\frac{8Z_i^2e^6  n_\sigma n_e n_i \ln\Lambda_{Ci}}{3 \sqrt{2\pi} (m c^2)^{3/2}(k_B T)^{1/2}}
            ,\ \ $$$${\rm [erg\ cm^{-3} s^{-1} Hz^{-1} Sr^{-1}]},
\end{equation}
and the absorption coefficient
\begin{equation}\label{Eq_abso_ff_Max_a}
\varkappa_{ff}^{M,\sigma}= \sum_i\frac{8Z_i^2e^6   n_e n_i \ln\Lambda_{Ci}}{3 \sqrt{2\pi}n_\sigma  c f^2 (m k_B T)^{3/2}}
      ,\quad {\rm [ cm^{-1}]},
\end{equation}
where
$e$ and $m$ are the electron charge and mass, $n_e$ is the number density of free electrons, $T$ is their temperature, $c$ is the speed of light, $k_B$ is the Boltzman constant, $n_\sigma$ is the refraction index of either ordinary or extraordinary wave mode, see Appendix\,\ref{S_appendix_disp}, 
the summation is performed over ions with the charge $Z_i$ and their number density $n_i$  in the ambient plasma.
These expressions obey Kirchhoff's law as needed for the thermal emissions.
Here, the Coulomb logarithm, $\ln\Lambda_{Ci}$, which originates in the classical treatment of the emission, can be expressed via the quantum mechanical 
\citet{1930RSPTA.229..163G} factor $G_i(T,f)$. The Gaunt factor depends on temperature $T$, frequency $f$, and the ion charge $Z_i$:
\begin{equation}
    \label{Eq_log2Gaunt}
    \ln\Lambda_{Ci}=\frac{3}{\sqrt{\pi}}G_i(T,f).
\end{equation}
Separating the factors that depend or not on the ion charge, we obtain the free-free emissivity and  absorption coefficient due to collisions of thermal electrons with positive ions in the arbitrarily ionized plasma (the e-ions contribution):
\begin{equation}\label{Eq_emis_ff_Max_z}
j_{f,ff}^{M,\sigma}=\frac{8e^6  n_\sigma  \ln\Lambda_C}{3 \sqrt{2\pi} (m c^2)^{3/2}(k_B T)^{1/2}}
n_e \sum_{i=1}^N g_i Z_i^2 n_i
            ,
\end{equation}

\begin{equation}\label{Eq_abso_ff_Max_z}
\varkappa_{ff}^{M,\sigma}=\frac{8e^6    \ln\Lambda_C}{3 \sqrt{2\pi}n_\sigma  c f^2 (m k_B T)^{3/2}} n_e\sum_{i=1}^N g_i Z_i^2 n_i
      ,
\end{equation}
where $\ln\Lambda_C=\ln\Lambda_{C1}$ is the Coulomb logarithm for singly-ionized ions; $g_i=G_i/G_1$ is the Gaunt factor for the ion with charge $Z_i$ normalized by that for the singly ionized ion $G_1$.

In natural plasmas, the Hydrogen is the most abundant element. For this case, we transform the above equations to a convenient form, where the contribution from all ions (but H) are described by a correction smaller than one.
Due to plasma neutrality, we have
\begin{equation}\label{Eq_n_e_ni}
 n_e=\sum_{i=1}^N Z_i n_i
      .
\end{equation}
Now we can express the sum in Eqns~(\ref{Eq_emis_ff_Max_z}) and (\ref{Eq_abso_ff_Max_z}) as the electron number density ($n_e$) and a correction as follows:

\begin{equation}\label{Eq_Z2_ni}
 \sum_{i=1}^N g_i Z_i^2 n_i=n_e+\sum_{i=1}^N g_i Z_i^2 n_i-\sum_{i=1}^N Z_i n_i
 = $$$$ \\ n_e\left(1+\sum_{i=2}^N Z_i(g_i Z_i-1) \frac{n_i}{n_e}\right)
 =\ \ 
  n_e\left(1+\zeta(T,f)\right)
      ,
\end{equation}

where
\begin{equation}\label{Eq_zeta3}
 \zeta(T,f)= \sum_{i=2}^N Z_i(g_i Z_i-1) \frac{n_i}{n_e} 
      .
\end{equation}
This expression depends on the temperature $T$ due to the temperature-dependent ionization and the Gaunt factor and on the frequency $f$ due to the Gaunt factor. This implies, that this value has to be tabulated as a function of two variables---$T$ and $f$.

Inserting Eqn.\,(\ref{Eq_Z2_ni}) into Eqns.\,(\ref{Eq_emis_ff_Max_z}) and (\ref{Eq_abso_ff_Max_z}):

\begin{equation}\label{Eq_emis_ff_Max_zeta}
j_{f,ff}^{M,\sigma}=
\frac{8e^6  n_\sigma  }{3 \sqrt{2\pi} (m c^2)^{3/2}}
\frac{n_e^2\ln\Lambda_C}{(k_B T)^{1/2}} \left(1+\zeta(T,f)\right)            ,
\end{equation}

\begin{equation}\label{Eq_abso_ff_Max_zeta}
\varkappa_{ff}^{M,\sigma}=\frac{8e^6   }{3 \sqrt{2\pi}n_\sigma  c f^2 (m )^{3/2}}
\frac{n_e^2\ln\Lambda_C}{(k_B T)^{3/2}} \left(1+\zeta(T,f)\right) 
      ,
\end{equation}
we obtain equations  suitable for the extension to the DEM treatment, which requires knowledge of the Coulomb logarithm and $\zeta$ function.  It is important to note that this treatment applies equally to plasma in ionization equilibrium as well as out of equilibrium, provided that the electron velocity distribution is Maxwellian.


\subsection{Treatment of the Gaunt factor}\label{GauntSection}


There is a controversy in the  semi-classical forms of the Coulomb logarithm (proportional to the Gaunt factor) in the literature. A consistent way of computing the Coulomb logarithm is via direct integration of the single-particle equations over the Maxwellian distribution.  The results of this treatment are given by \citet{1997riap.book.....Z} (Section 11.1). The low-temperature asymptote is
\begin{equation}
\label{Eq_Clog_low}
 \ln\Lambda_C=\frac{\pi}{\sqrt{3}} G_i(T,f)= \ln\left[\frac{(2k_B T)^{3/2}}{\pi \delta^{5/2}e^2 Z_i m^{1/2}f} \right],
\end{equation}
while the high-temperature one is
\begin{equation}
\label{Eq_Clog_high}
 \ln\Lambda_C=\frac{\pi}{\sqrt{3}} G_i(T,f)= \ln\left[\frac{4k_B T}{\delta\, h\, f} \right],
\end{equation}
where $\delta=e^C$, $C\approx 0.577$ is the Euler constant, and $h$ is the Planck constant. The low-temperature asymptote depends on $Z_i$; thus,  the boundary temperature separating the asymptotic regimes depends on $Z_i$ as well. For $Z_i=1$, the boundary temperature, where the asymptotes intersect, is $T_*\approx 0.89125$\,MK. Often, numeric equations equivalent to (\ref{Eq_Clog_low}) and (\ref{Eq_Clog_high}) are used, where all constants are substituted by their numerical values, which yields:
\begin{equation}
\label{Eq_Clog_num}
 \ln\Lambda_C=\left\{
 \begin{array}{ll}
 17.718414+\ln (T^{3/2}/Z_i) - \ln f, \quad T<T_*
    \\
 24.569056 +\ln T - \ln f, \quad T>T_*.
  \end{array}
  \right.
\end{equation}

It is worthwhile to note that there are other forms of the Coulomb logarithm. In particular, \citet{1985ARA&A..23..169D} used an approximation of a single-particle value with the thermal velocity $v=\sqrt{k_BT/m}$, rather than integration over velocities, for $\ln\Lambda_C$, which yielded different equations (Eqns 18 and 19 in his paper) for $\ln\Lambda_C$ compared with Eqns.\,(\ref{Eq_Clog_low}) and (\ref{Eq_Clog_high}). This simplified \citet{1985ARA&A..23..169D} approach results in the terms 17.54 (instead of 17.718414) and 24.5 (instead of 24.569056) in Eqn.\,(\ref{Eq_Clog_num}). These exact and approximate numbers are consistent with each other. However, in the numerical estimate, \citet{1985ARA&A..23..169D} used 18.2 (instead of 17.54), which was noted and criticized by \citet{2016SSRv..200....1W}, while the boundary temperature was incorrectly assigned to 0.2\,MK (instead of 0.89125\,MK). Those wrong numbers propagated to many publications and textbooks including
\citet{FT_2013}. {This introduces an} 
error of $\sim3\%$. 

Here, we employ the treatment of the Gaunt factor by \citet{2014MNRAS.444..420V} for nonrelativistic plasma  ($T\ll5\times10^9$\,K), which is an extension of results by  \citet{1961ApJS....6..167K}.
The Gaunt factors were computed for a point charge, so they are explicitly valid for the corresponding bare nuclei
(fully ionized element), but widely used for other ions with the same charge.
The approximation that the Gaunt factor is the same for each ion with the same charge regardless of the element is a rather good one. Indeed, far collisions are mainly responsible for radio emission; thus,
the approximation of any ion as a point charge is a rather good one in the radio domain even for ions that are not fully ionized.

We do not consider the relativistic case \citep{2015MNRAS.449.2112V} for the following reasons:
\begin{itemize}
    \item That hot (relativistic) plasma is not expected in the solar corona.
    \item Available hydro codes \citep[e.g., EBTEL;][]{2008ApJ...682.1351K,2012ApJ...752..161C,2012ApJ...758....5C}  do not typically provide DEM at temperatures above a few$\times10^8$\,K. 
    \item At the relativistic regime, the non-dipole emission due to electron-electron collisions will become non-negligible, which is anyway ignored.
\end{itemize}

\begin{figure*}
  \centering
    \includegraphics[width=\textwidth]{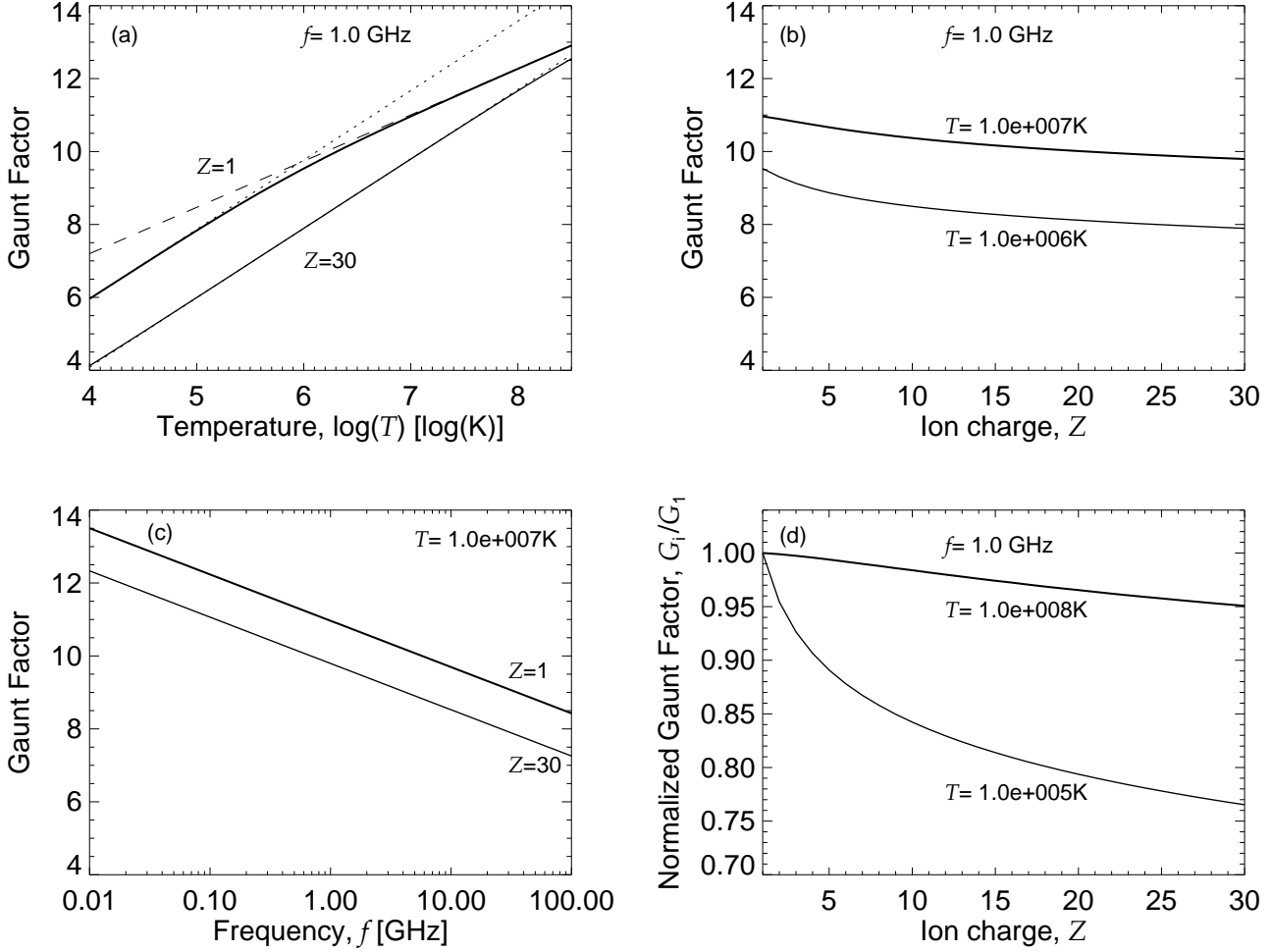}
  \caption{Dependence of the Gaunt factor on the involved parameters:  plasma temperature (a), ion charge (b), and frequency (c). Panel  (a) also shows the high- and low-temperature asymptotes (\protect\ref{Eq_Clog_num}) by dashed and dotted lines, respectively. Panel  (d) shows the Gaunt factor for a given ion $Z_i$ normalized by that for the proton Gaunt factor $G_1$: $g_i=G_i/G_1$.}\label{f_gaunt}
\end{figure*}

Figure\,\ref{f_gaunt} shows  the  Gaunt factor, tabulated by \citet{2014MNRAS.444..420V}, as a function of the temperature $T=10^4-3.16\times10^8$\,K on panel (a), of the ion charge $Z=1-30$ on panel (b), and of the frequency $f=0.01-100$\,GHz on panel (c). Panel  (d) shows an example of the Gaunt factor normalized by that for a proton.

Figure\,\ref{f_gaunt}a also shows the asymptotes of the Gaunt factor obtained using Eqn.\,(\ref{Eq_Clog_num}) for $Z=1$ and $Z=30$ by the dashed and dotted lines.
Direct comparison between the dashed/dotted and solid curves shows that the asymptotic expressions work well everywhere, except, perhaps, a narrow region around their intersection.
Although the asymptotic expressions might be sufficient for most practical  applications, here we employ the most accurate treatment by \citet{2014MNRAS.444..420V}.


\subsection{Contribution of highly ionized ions}

In the hot corona, many ions are in high ionization states. Given that the free-free emissivity and absorption coefficient are proportional to $Z_i^2$, where $Z_i$ is the ion charge, even relatively minor ions, such as O, Fe, C can give a significant contribution to the free-free emission; $\gtrsim$1--3\% each.

In Eqns.\,(\ref{Eq_emis_ff_Max_zeta}) and (\ref{Eq_abso_ff_Max_zeta}) the cumulative effect of all ions with $Z_i\geq2$ is included in the term $\zeta(T,f)$, where (a rather weak) dependence on frequency $f$ appears due to the Gaunt factor. The $\zeta(T,f)$ function depends both on the elemental abundances and on each element's ionization state, which are different in equilibrium and non-equilibrium conditions. In this work, we will provide the $\zeta(T,f)$ function calculated under the assumption of ionization equilibrium using the CHIANTI spectral code \citep{1997A&AS..125..149D, 2019ApJS..241...22D}, as detailed below. In this framework 
it is straightforward to users to replace the CHIANTI values for the equilibrium ion fractions by their own values calculated under non-equilibrium conditions, depending on the physical process they model.  We carried out the calculations for the plasma element abundances typical of the solar corona \citep{1992PhyS...46..202F}, as well as for two abundance models obtained for the solar photosphere \citep{2011SoPh..268..255C, 2015A&A...573A..25S}. 
The main difference between these models is that in the solar corona all elements with low First Ionization Potential (FIP) are overabundant by a factor of about four over their photospheric abundances.  However, this overabundance factor is somewhat uncertain---while it has been extensively used in the field, its value has been suggested to be variable with time \citep{2001ApJ...555..426W} 
and also within the same active region; also, the absolute correction of the abundances has been disputed, as some studies suggest that the FIP effect affects both low-FIP and high-FIP ions \citep{2012ApJ...755...33S}. Furthermore, solar abundances can be different from those of other stars, both in the photosphere and in the corona, even to the point that some stars exhibit an inverse FIP effect \citep{2015LRSP...12....2L}. 
So, the software we implemented to carry out the calculations allows the user to interactively select an abundance data set that is best suited for the star or the solar region under consideration. 


In order to utilize the data in the CHIANTI spectral code, the dependence of the $\zeta(T,f)$ on the plasma elemental and ionic composition was made explicit, by developing the $n_i/n_e$ ratio as
\begin{equation}
    \frac{n_i}{n_e} = \frac{n{\left({X_i}\right)}}{n{\left({X}\right)}}\frac{n{\left({X}\right)}}{n{\left({H}\right)}}\frac{n{\left({H}\right)}}{n_e}
\end{equation}

\noindent
where $n{\left({X_i}\right)}/n{\left({X}\right)}$ is the fraction of the element $X$ in the $i$ ionization stage, $n{\left({X}\right)}/n{\left({H}\right)}$ is the abundance of element $X$ relative to the abundance of $H$, and  $n{\left({H}\right)}/n_e$ is the ratio between Hydrogen (regardless of its ionization stage) to the density of the free electrons. Using this expression, the $\zeta(T,f)$ function can be rewritten as a double sum over both elements ($i$) and ions ($m$):

\begin{equation}
    \zeta(T,f)= $$$$\sum_{i=2}^N \sum_{m=2}^{i}{\left[{Z_m(g_m{\left({T,f}\right)}Z_m - 1) \frac{n{\left({X_{i,m}}\right)}}{n{\left({X_i}\right)}}\frac{n{\left({X_i}\right)}}{n{\left({H}\right)}}\frac{n{\left({H}\right)}}{n_e}}\right]},
\end{equation}

\noindent
where the summation extends to all elements from He ($i=2$) to Zn ($i=30$), and to all ions with ($Z_m\geq2$) including the bare nuclei ($m=i$); here, clearly, the charge $Z_m$ of the ion with $m$ electrons removed from the atom is $Z_m=m$. By making the dependence on the ion fraction $n{\left({X_{i,m}}\right)}/n{\left({X_i}\right)}$ and on the element abundance $n{\left({X_i}\right)}/n{\left({H}\right)}$ explicit, users can easily plug in their values that better represent the plasma they are considering.

\begin{figure*}
  \centering
  \includegraphics[width=\textwidth]{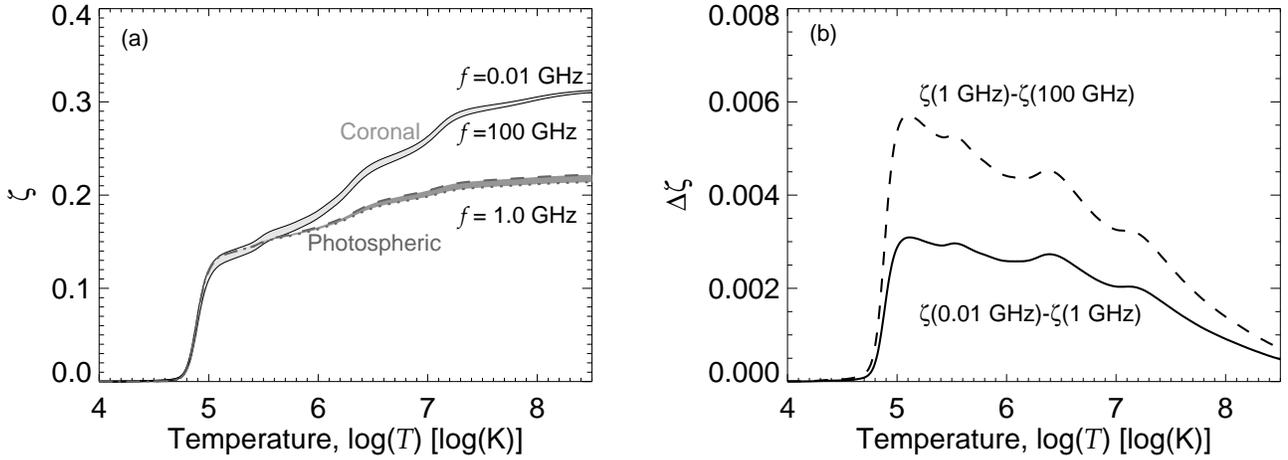}
  \caption{(a) Dependence of the $\zeta$ function on temperature computed from CHIANTI code for frequencies between 0.01 and 100\,GHz is shown by a light gray shadow bounded by solid curves. The ion abundances come from the CHIANTI database, and the elemental abundances from \protect\citet{1992PhyS...46..202F}. The elements included in the calculation are  He to Zn (Z=2-30). For comparison, the dark gray segment shows the $\zeta$ function at $f=1$\,GHz for photospheric abundances within the range reported by \protect\citet{2011SoPh..268..255C} (dashed line) and  \protect\citet{2015A&A...573A..25S} (dotted line). (b) Difference between $\zeta$ function at 1\,GHz and 0.01 or 100\,GHz, respectively for the coronal abundance.} \label{f_zeta}
\end{figure*}

As an example, for the assumptions adopted above, we tabulated $\zeta(T,f)$ function for a range of frequencies between 10\,MHz and 100\,GHz. 
Figure\,\ref{f_zeta} shows this function in a broad range of temperatures, where  $\zeta(T,f)$ varies between 0 and $\sim0.3$. This is consistent with an intuitive expectation that at low temperatures the atoms are at most singly ionized, while there are mostly bare nuclei at the highest temperatures.
The difference between the photospheric and coronal cases is quite significant and caused by the coronal low-FIP element enhancement; primarily, iron ions. This is due to the fact that the most abundant high-FIP ions, namely He, C, N, O, Ne, have very similar abundance values in both data sets, and at temperatures larger than 1~MK have very few ions surviving under equilibrium conditions; on the contrary, at those temperatures the most abundant low-FIP elements (Mg, Si, Fe) are distributed among many stages of ionization, providing a larger number of terms to the $\zeta$ function. The dependence of the $\zeta$ function on the emission frequency is very weak.

\begin{figure}
  \centering
  \includegraphics[width=0.5\textwidth]{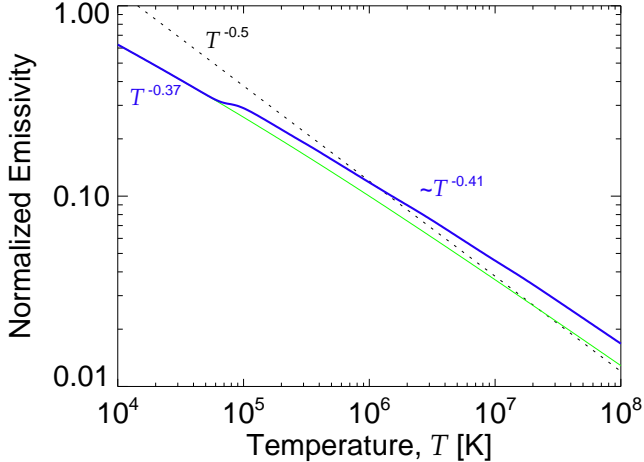}
  \caption{Dependence of the free-free emission emissivity at 1\,GHz on temperature: green line shows the result for a hydrogen plasma with the account of exact form of the Gaunt factor, while the blue line also accounts for the $\zeta$ function computed for coronal abundances in ionization equilibrium. For comparison, the dotted line shows the $T^{-0.5}$ dependence, cf. Eqn.\,(\ref{Eq_emis_ff_Max_z}). } \label{f_j_vs_T}
\end{figure}

{Figure\,\ref{f_j_vs_T} illustrates the effect of the plasma composition and ionization on the dependence of the e-ions free-free emissivity on the plasma temperature. Note that dependence of the logarithmic Gaunt factor on the temperature is often neglected and the dependence   $j_{ff} \propto T^{-1/2}$ is used \citep[cf.][]{2001ApJ...561..396Z}, {while in other cases an approximation such as a power law is used}  \citep[e.g., $j_{ff} \propto T^{-0.35}$, ][]{2003ApJ...589.1054L}. In contrast, Figure\,\ref{f_j_vs_T} shows that the true emissivity does not follow any power-law. At low temperatures 
the dependence is approximately a power-law with an index $-0.37$. Then, between 60\,kK and 100\,kK the emissivity displays a plateau, where the term $\propto T^{-1/2}$ is almost balanced by the increase of the $\zeta$ factor due to the second ionization of the singly ionized He ions. At higher energies the emissivity does not follow a power law, but can approximately be represented as $j_{ff} \propto T^{-0.41}$.
}

\begin{figure}
\resizebox{8.5cm}{!}{\includegraphics{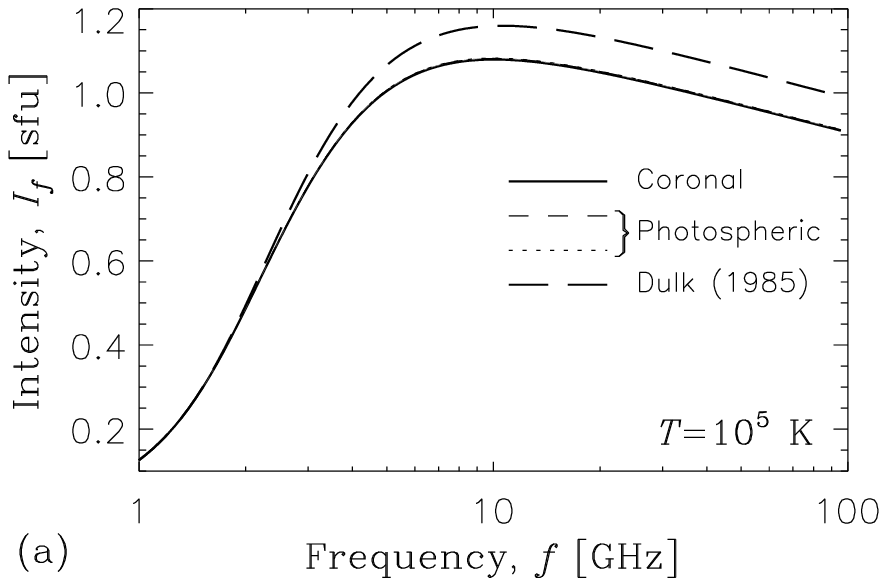}}\\
\resizebox{8.5cm}{!}{\includegraphics{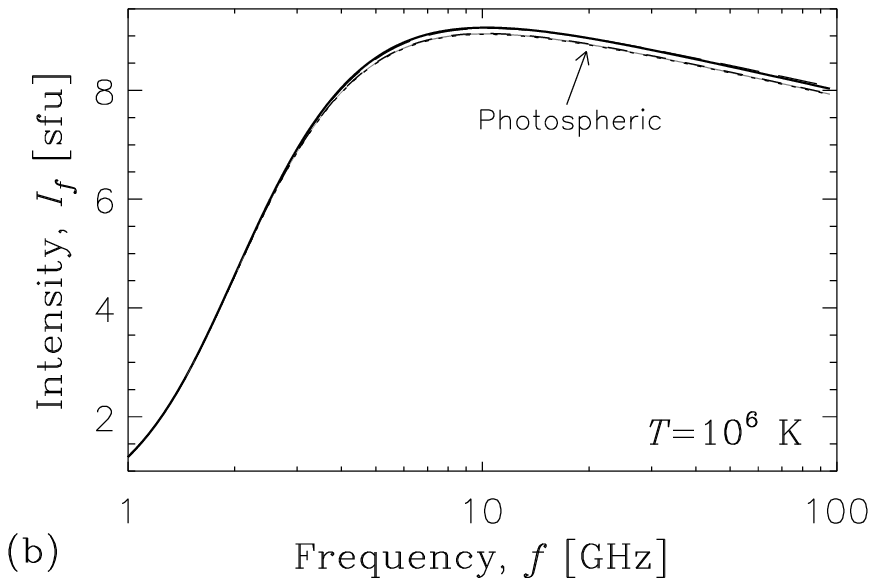}}\\
\resizebox{8.5cm}{!}{\includegraphics{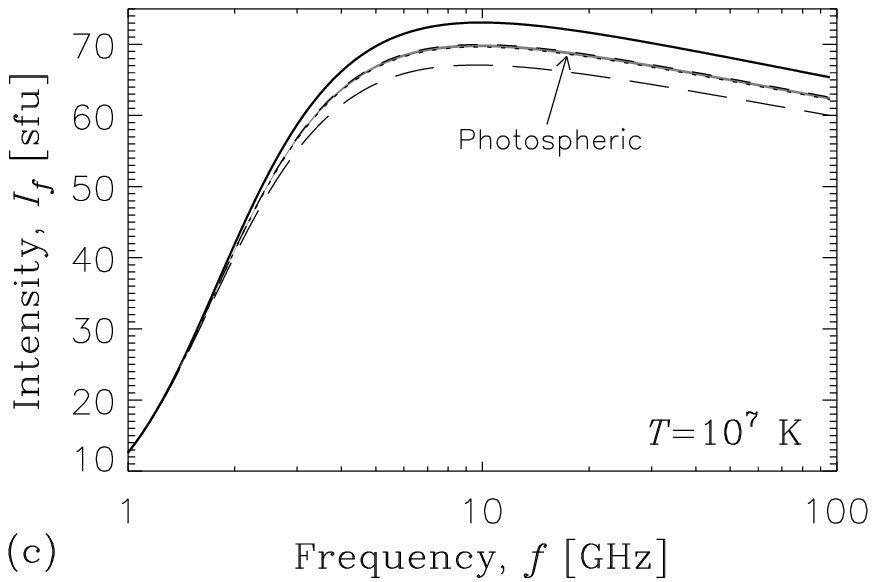}}
\caption{ Effect of abundance 
on the e-ions component of the free-free emission. The spectra are computed for isothermal plasma and homogeneous source, using the new code with $\zeta$-function and exact Coulomb logarithm. Solid lines correspond to the coronal abundance by \protect\citet{1992PhyS...46..202F}, and the dark gray stripes correspond to the photospheric abundances within the range reported by \protect\citet{2011SoPh..268..255C} (dashed lines) and \protect\citet{2015A&A...573A..25S} (dotted lines). For comparison, the spectra computed according to the formulae of \protect\citet{1985ARA&A..23..169D} are plotted, too (with long-dashed lines). 
Simulation parameters: source area $S = 10^{20}$ $\textrm{cm}^2$, electron number density $n_e = 10^9$ $\textrm{cm}^{-3}$, no magnetic field. Three different temperatures are considered; the source depths are: $L = 2\times 10^9$ cm for $T = 10^5$ K, $L = 4\times 10^{10}$ cm for $T = 10^6$ K, and $L= 8\times 10^{11}$ cm for $T = 10^7$ K; the depths are chosen to provide the spectral peak always at about 10 GHz.}\label{FFspectra}
\end{figure}

Figure \ref{FFspectra} illustrates the effect of the heavy element abundances and their ionization states at various high temperatures on the free-free emission {(at these high temperatures, there are no neutral atoms; thus, only the e-ions contribution is considered; the electron-neutral collisions become important for a colder plasma; see Section \ref{NeutralsSection})}. As expected, at relatively low temperatures ($10^5$\,K) both  coronal and photospheric abundance models provide nearly the same free-free radio emission spectra,  while at higher temperatures, the plasma with the coronal abundances provides a measurably higher emission intensity in the optically thin range than the plasma with the photospheric abundances does for the same reason discussed above for the $\eta$ function; the difference is about 5\% at the temperature of $\sim 10^7$ K. For comparison, we plot the emission spectra obtained using the approximate formulae by \citet{1985ARA&A..23..169D}  for a hydrogen-helium plasma with typical solar abundance and the (incorrect) expressions for the Coulomb logarithm discussed in Section \ref{GauntSection}. Although these formulae provide a surprisingly good agreement with the more accurate treatment (for the coronal abundances) for $T\simeq 10^6$ K, they overestimate the emission intensity at lower temperatures (by up to $\sim 20\%$ at $T\simeq 10^5$ K) and underestimate it at higher temperatures (by up to $\sim 10\%$ at $T\simeq 10^7$ K and the coronal abundance model).

\subsection{Contribution of neutrals}\label{NeutralsSection}


In contrast to the corona, the chromosphere is relatively cool, so there are numerous neutral atoms, primarily H and He. \citet{1974ApJ...187..179S} reported an accurate expression for the absorption coefficient due to collisions of thermal electrons with neutral hydrogen, often called the \textit{$H^-$ absorption}, which we reproduce here with a correction for the refraction index $n_\sigma$:

\begin{equation}
\label{Eq_Hminus_1}
\varkappa_{ff}^{\sigma,H-}=\frac{128\pi^{3/2}}{3n_\sigma} n_e n_{HI} k_B T \frac{\alpha a_0^5}{E_H}\left(\frac{E_H}{hf}\right)^2 \frac{Q}{k_T}
,
\end{equation}
where $n_{HI}$ is the number density of neutral hydrogen, $\alpha=e^2/(\hbar c)$ is the fine structure constant, $a_0=\hbar^2/(me^2)$ is the Bohr radius, $E_H=e^2/(2a_0)$ is the ionization energy of hydrogen, $k_T=(k_B T/E_H)^{1/2}$, $Q$ is the temperature-dependent electron-hydrogen momentum transfer cross-section. \citet{1974ApJ...187..179S} provided an analytical approximation of the cross-section $Q$ in the form $Q=Q_0\exp(-\xi)$, where $Q_0=65$ and

\begin{equation}
\label{Eq_xi_H}
\xi=4.862k_T(1 - 0.2096\,k_T+ 0.0170\,k_T^2 - 0.00968\,k_T^3);
\end{equation}
this approximation is valid in the range of temperatures of $2500\lesssim T\lesssim 50\,000$ K.

Substituting some of the constants in Eq.\,(\ref{Eq_Hminus_1}), \citet{1974ApJ...187..179S} reduces it to the form

\begin{equation}
\label{Eq_Hminus_2}
\varkappa_{ff}^{\sigma,H-}=
2.1451976\times10^{-29}
\frac{ n_e n_{HI} k_B T}{n_\sigma} \left(\frac{E_H}{hf}\right)^2 \frac{\exp(-\xi)}{k_T}
,
\end{equation}
which can be further simplified as follows
\begin{equation}
\label{Eq_Hminus_3}
\varkappa_{ff}^{\sigma,H-}=
1.0840\times10^{-3}\
\frac{ n_e n_{HI} (k_B T)^{1/2}}{n_\sigma f^2}  \exp(-\xi)
.
\end{equation}

\cite{1974A&A....30..293S} provided a similar treatment for atoms of several noble gases, of which we only explore the results for helium, because even the contribution to the opacity from the most abundant neutral helium does not exceed a few percent. Assuming $hf\ll 2k_B T$, which is always valid for solar atmosphere temperatures in the radio domain, and adding a correction for the refraction index, Eqs.\,(7) and (11) from \citet{1974A&A....30..293S} reduce to

\begin{equation}
\label{Eq_He_minus_1}
\varkappa_{ff}^{\sigma,He-}
=\frac{ n_e n_{HeI} k_B T}{n_\sigma}
\left(\frac{E_H}{hf}\right)^2 \xi_{He}(T)
,
\end{equation}
where the function $\xi_{He}(T)$ can be approximated (in the range of temperatures of $2500\lesssim T\lesssim 25\,000$ K) by an analytical expression
\begin{equation}
\label{Eq_xi_He}
\xi_{He}=\frac{10^{-30}}{k_T}(1.868 + 7.415 k_T -22.56 k_T^2 + 15.59 k_T^3).
\end{equation}


The corresponding emissivities ($j$) are linked to the absorption coefficients ($\varkappa$) by the Kirchhoff law \citep[see, e.g.,][]{FT_2013}

\begin{equation}
\label{Eq_emis_via_abs}
j=\frac{n_\sigma^2 f^2  k_B T}{c^2}\varkappa.
\end{equation}

It is interesting to note that these emissivities are not dependent on frequency. In the chromosphere, where the collisions with neutrals are important, the concept of the DEM is not typically used. In fact, it would be difficult to consistently use this concept, because the emissivity and absorption coefficient depend on a product $n_e n_{HI}$ or $n_e n_{HeI}$, rather than on $n_e^2$, and it is difficult to express one via the other. Often, chromospheric models \citep[see, e.g.,][]{2009ApJ...707..482F, 2014SoPh..289..515F} supply the user with number densities of the neutral and ionized hydrogen, which might deviate from the local thermodynamic equilibrium (non-LTE). If such non-LTE number densities are provided, our codes explicitly take them into account to compute free-free emission. If only the total number density is provided, the code will use the Saha equation to compute equilibrium ionization of hydrogen for the temperatures below 0.1\,MK. 
In contrast,  non-LTE helium ionization is typically not provided by models; thus, to compute helium ionization states at low temperatures, we use Saha equation. At low temperatures, when the Saha equation predicts literally zero ionization fraction, our code assumes the ionization fraction of $10^{-3}$ due to easily ionized metals, which ensures a non-zero free-free emission even for those relatively low temperatures.

\begin{figure*}
\resizebox{8.5cm}{!}{\includegraphics{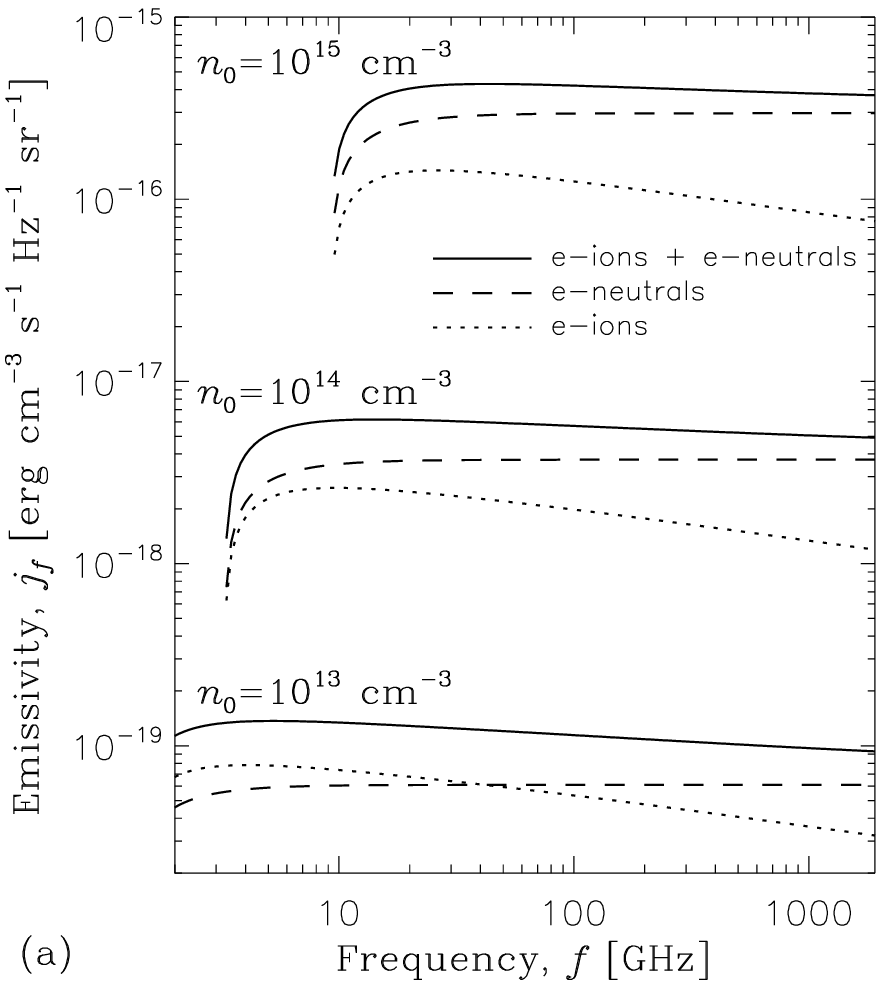}}%
\resizebox{8.5cm}{!}{\includegraphics{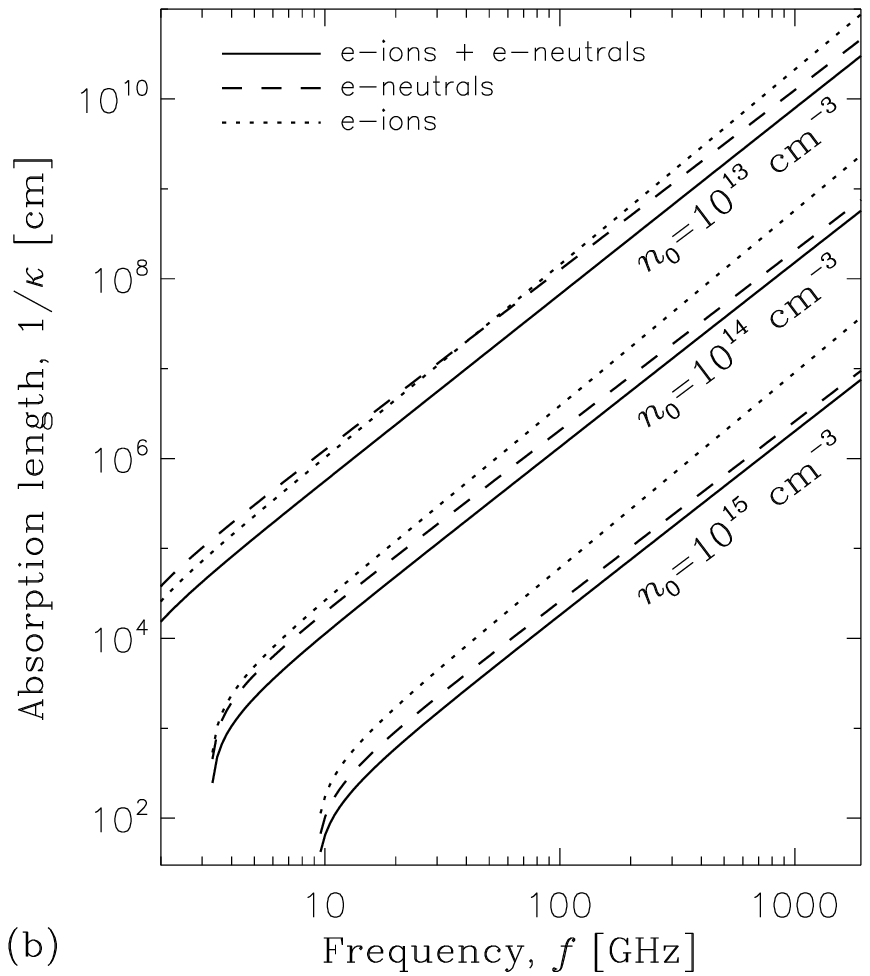}}
\caption{{The free-free plasma radio emissivities (a) and absorption lengths (b) in a cool plasma with chromospheric parameters. The e-ions and e-neutrals (H$^-$ and He$^-$) contributions are shown by the dotted and dashed lines, respectively, while the total free-free emissivities are shown by solid lines. The emissivities are computed for isothermal plasma using the new code with $\zeta$-function \protect\citep[photospheric abundance model by][]{2011SoPh..268..255C} and exact Coulomb logarithm. Simulation parameters: plasma temperature $T=5000$ K, no magnetic field ($j_f\equiv j_f^X\equiv j_f^O$). Three different gas densities $n_0$ are considered; the electron and neutral number densities are computed using the Saha equation, which provides  $n_{\mathrm{e}}\simeq 2.15\times 10^{10}$ $\textrm{cm}^{-3}$, $n_{\mathrm{H}} \simeq 9.21\times 10^{12}$ $\textrm{cm}^{-3}$, and $n_{\mathrm{He}}\simeq 7.8\times 10^{11}$ $\textrm{cm}^{-3}$ for $n_0=10^{13}$ $\textrm{cm}^{-3}$, $n_{\mathrm{e}}\simeq 1.31\times 10^{11}$ $\textrm{cm}^{-3}$, $n_{\mathrm{H}}\simeq 9.22\times 10^{13}$ $\textrm{cm}^{-3}$, and $n_{\mathrm{He}}\simeq 7.8\times 10^{12}$ $\textrm{cm}^{-3}$ for $n_0=10^{14}$ $\textrm{cm}^{-3}$, and $n_{\mathrm{e}}\simeq 1.05\times 10^{12}$ $\textrm{cm}^{-3}$, $n_{\mathrm{H}}\simeq 9.22\times 10^{14}$ $\textrm{cm}^{-3}$, and $n_{\mathrm{He}}\simeq 7.8\times 10^{13}$ $\textrm{cm}^{-3}$ for $n_0=10^{15}$ $\textrm{cm}^{-3}$.}}\label{HminusSpectra}
\end{figure*}

Figure \ref{HminusSpectra} shows {the free-free emissivity and characteristic absorption length
due to collisions of thermal electrons with ions (the  e-ions component) and neutrals (the e-neutrals, H$^-$ and He$^-$ components). These quantities are computed for a range of parameters typical for the solar chromosphere \citep[see, e.g.,][and references therein]{2017A&A...601A..43L}.
For the considered here temperature of 5,000\,K, relative contributions of these two components depend on the total Hydrogen number density $n_0$. For a relatively low density, $n_0=10^{13}$\,cm$^{-3}$, the contributions are comparable to each other; the e-ions contribution dominates at lower frequencies, while the e-neutrals one at the higher frequency, because the latter does not decrease with frequency unlike the former one. For the higher densities, the e-neutral contribution dominates, because the ionization fraction is relatively low under the considered conditions. The right panel indicates that both contributions to the effective absorption length are important and have to be taken into account when the plasma temperature is in the range of a few thousand K.
}

\subsection{Effect of ambient magnetic field on the free-free emission}

So far, we have not explicitly considered the effect of the ambient magnetic field of the free-free opacity, other than accounting for the refractive index $n_\sigma$, which differs for the ordinary and extraordinary wave modes (see Appendix\,\ref{S_appendix_disp}). The expressions discussed above have the form $\varkappa_{ff}^\sigma=\varkappa_{ff, 0}/n_\sigma$, where $\varkappa_{ff, 0}$ is the free-free absorption coefficient for $B=0$ and $n_\sigma=1$. This approach is valid at frequencies much larger than the gyro frequency. In stellar coronae the gyro frequencies are not necessarily small enough to justify this simplified treatment, so we employ a more accurate treatment of \citet[][Eqn. 14]{1968SvA....12..245Z}\footnote{Note that Eqn. (17) in \citet{1968SvA....12..245Z} contains a typo.} 
\begin{equation}
\label{Eq_kappa_ff_B}
\varkappa_{ff}^\sigma=\frac{\varkappa_{ff, 0}}{n_{\sigma}}F_\sigma
\end{equation}
where 
\begin{equation}
\label{Eq_F_sigma}
 F_\sigma = 2~\frac{
 \sigma\sqrt{\mathcal{D}}\left[u\sin^2\theta+2(1-v)^2\right] - u^2\sin^4\theta }
 {\sigma\sqrt{\mathcal{D}}\left[2(1-v)-u\sin^2\theta+\sigma\sqrt{\mathcal{D}}\right]^2}
 ,
\end{equation}
\begin{equation}
\mathcal{D}=u^2\sin^4\theta+4u(1-v)^2\cos^2\theta,
\end{equation}
\begin{equation}
u=\left(\frac{f_{Be}}{f}\right)^2,\qquad
v=\left(\frac{f_{\mathrm{pe}}}{f}\right)^2,
\end{equation}
$f_{Be}=eB/(2\pi mc)$ is the gyro frequency, $f_{\mathrm{pe}}=e\sqrt{n_{e}/(\pi m_{\mathrm{e}})}$ is the electron plasma
frequency, $\theta$ is the angle between the line of sight (LOS) and the magnetic field, {$\sigma=\pm1$ for O(X) modes}. According to Kirchhoff's law, Eq.\,(\ref{Eq_emis_via_abs}), the same $F_\sigma$ factor has to be applied to the emissivity. 

\begin{figure*}
  \centering
  \includegraphics[width=\textwidth]{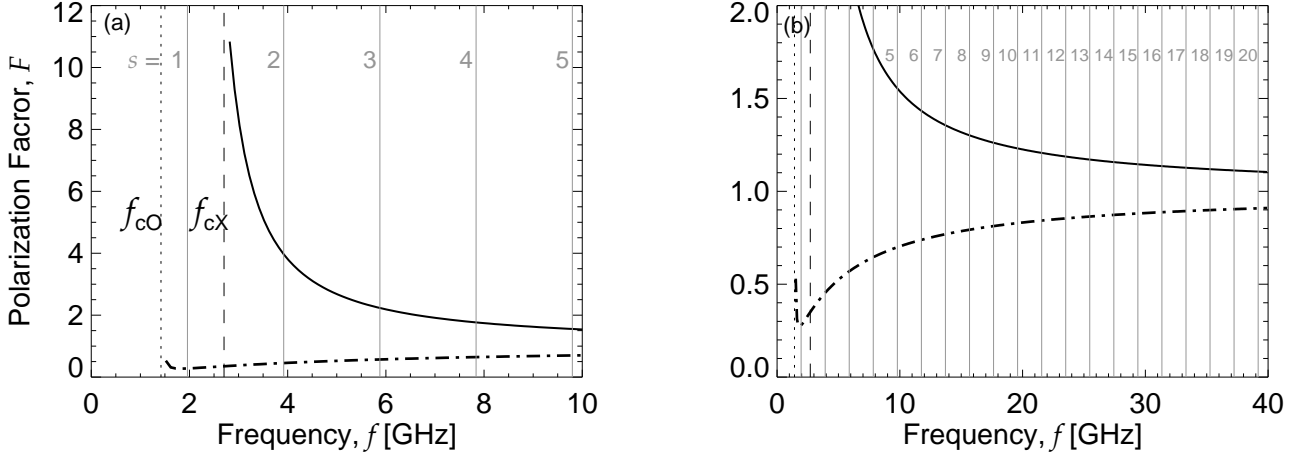}
  \caption{(a) Dependence of the polarization factor $F_\sigma$ on frequency $f$ for the extraordinary (thick solid line) and ordinary (thick dash-dotted line) in the spectral range covering five lowest gyro harmonics for the following parameters: $n_e=2.5\times10^{10}$\,cm$^{-3}$ ($f_{\mathrm{pe}}\approx1.42$\,GHz), $B=700$\,G ($f_{Be}\approx1.96$\,GHz), and $\theta=10^\circ$. 
  The cutoff frequencies for the ordinary wave, $f_{cO}$, and extraordinary wave, $f_{cX}$, are shown by the vertical dotted and dashed lines, respectively. Thin vertical solid lines with the numbers $s$ next to them indicate integer multiples $sf_{Be}$ of the gyro frequency $f_{Be}$.  (b) The same functions over a narrower $y$-range, while extended range of the frequencies.
  }\label{f_zlotnik}
\end{figure*}

Figure\,\ref{f_zlotnik} gives an example of the $F_\sigma$ factors computed for  modest magnetic field and electron density. While this factor does not change the ordinary mode opacity by more than 50\%, for the extraordinary mode it can be as large as $\sim10$. {The emissivity and absorption of the X-mode increase because electrons rotate in the magnetic field in the same direction as the electric field vector in the X-mode waves resulting in a stronger coupling between them compared with the case of zero magnetic field. } That large enhancement of the free-free opacity takes place at the range of lowest harmonics ($s\leq 3$) of the gyro frequency, where the total opacity of the plasma is typically dominated by the GR process considered in the next section. However, as follows from Eq.\,(\ref{Eq_tau_Max_a}), the GR opacity contains a factor $\sin^{2s-2}\theta$, and thus, it is zero along the magnetic field. In fact, there is a finite GR `transparency window' around this direction, where the free-free opacity dominates. Figure\,\ref{f_zlotnik} further shows that the polarization factors are noticeably different from the asymptotic value 1 up to rather high frequencies even for the adopted moderate value of the magnetic field $B=700$\,G. Thus, the account of this factor is important in the radio domain. In the modeling of the free-free component we can safely apply this factor to the total free-free opacity (and the same factor to the emissivity) of the given voxel.

\begin{figure}
\resizebox{8.5cm}{!}{\includegraphics{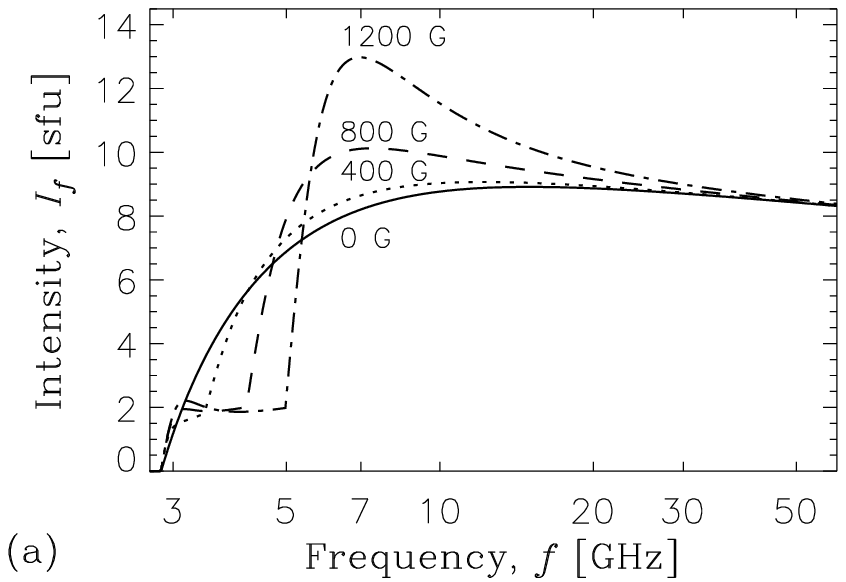}}\\
\resizebox{8.5cm}{!}{\includegraphics{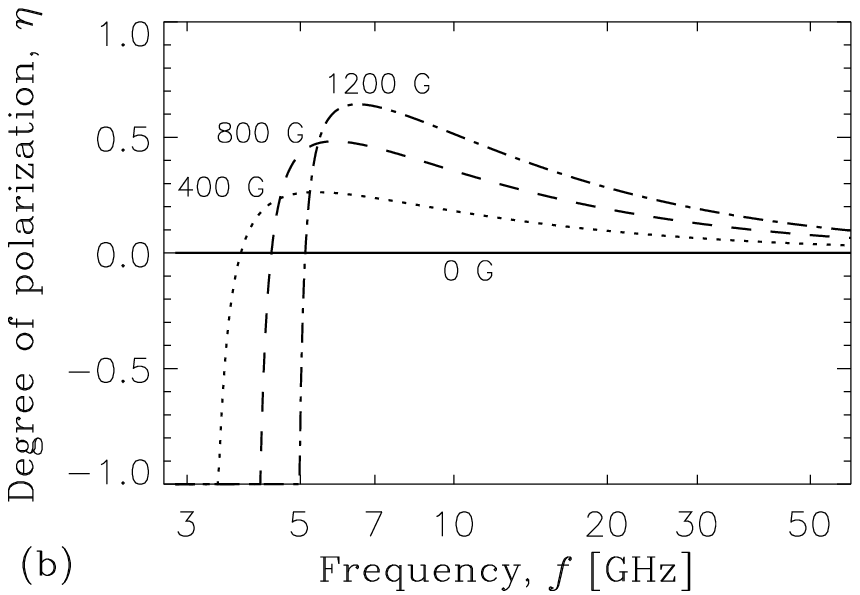}}\\
\resizebox{8.5cm}{!}{\includegraphics{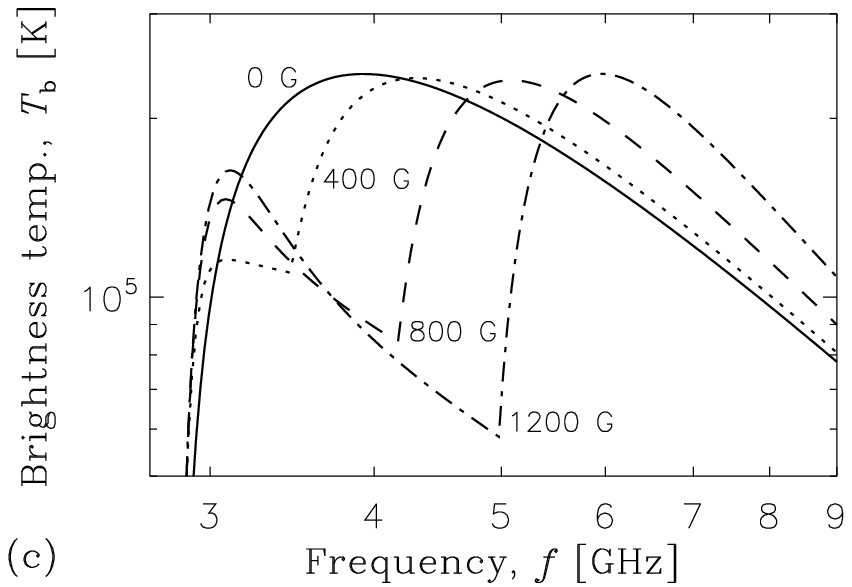}}
\caption{Effect of the magnetic field on the free-free emission. The spectra are computed for isothermal plasma and homogeneous source, using the new code with $\zeta$-function (coronal abundance model) and exact Coulomb logarithm. 
Simulation parameters: source area $S = 10^{20}$ $\textrm{cm}^2$, source depth {$L = 4\times 10^6$ cm}, plasma temperature $T = 10^6$ K, the electron number density {$n_e = 10^{11}$ $\textrm{cm}^{-3}$}, the viewing angle {$\theta = 30^{\circ}$}; four different magnetic field values are indicated in the panels. Note that in the brightness temperature plot (panel (c)), the frequency range is different from that in two other panels.}\label{FF_MFeffect}
\end{figure}

Figure \ref{FF_MFeffect} shows the free-free emission in the presence of ambient magnetic field. The increase of magnetic field strengths results in a noticeable increase of the free-free opacity of the X-mode; thus, the transition from the optically thick to thin regime occurs at a higher frequency. Therefore, the
total  intensity of the free-free emission goes up above the X-mode cutoff frequency $f_{cX}$ (see Appendix\,\ref{S_appendix_disp}), where 
the degree of polarization can be rather large in the X-mode sense. At the frequencies below the cutoff frequency $f<f_{cX}$, the X-mode cannot propagate; thus, the free-free emission is 100\% O-mode polarized. This is why the total brightness at these low frequencies is only one half of the free-free emission brightness without magnetic field. 

\section{Gyroresonance emission from Maxwellian distributions}

Classical theory of the GR radiation from a Maxwellian plasma \citep{1962SvA.....6....3Z, 1962ApJ...136..975K, 1997riap.book.....Z} yields the optical depth of the $s$-th gyro layer---a narrow surface where a given frequency matches a small integer multiple $s$ of the gyro frequency $f_{Be}=eB/(2\pi mc)$---by integrating the absorption coefficient along the line of sight \citep[for the derivation and notations, see][and Appendix\,\ref{S_appendix_disp}]{Fl_Kuzn_2014}:

\begin{equation}\label{Eq_tau_Max_a}
 \tau_{s}^{M,\sigma}=\int\limits_{-\infty}^{\infty}\varkappa_{s}^{M,\sigma}(z) dz=$$$$
        \frac{\pi e^2n_e }{f mc}
       \left(\frac{k_B T}{mc^2}\right)^{s-1}
       \frac{s^{2s}n_\sigma^{2s-3}\sin^{2s-2}\theta}{2^{s-1} s!(1+T_\sigma^2) } L_B
       [T_\sigma\cos\theta + L_\sigma\sin\theta+1]^2 
             ,
\end{equation}
and, accordingly, the emissivity along the line of sight:
\begin{equation}\label{Eq_emisZ_Max_a}
 J_{f,s}^{M,\sigma}=\int\limits_{-\infty}^{\infty}j_{f,s}^{M,\sigma} (z) dz=$$$$
        \frac{\pi e^2n_e f}{c}
       \left(\frac{k_B T}{mc^2}\right)^{s}
       \frac{s^{2s}n_\sigma^{2s-1}\sin^{2s-2}\theta}{2^{s-1} s!(1+T_\sigma^2) } L_B
       [T_\sigma\cos\theta + L_\sigma\sin\theta+1]^2 
             ,
\end{equation}
where 
$T_\sigma$ and $L_\sigma$ are the components of the polarization ellipse of either ordinary or extraordinary wave mode (Appendix\,\ref{S_appendix_disp}), 
$s$ is an integer number of gyro harmonics. Here, the spatially nonuniform magnetic field has been expanded into a series in the gyro layer and its vicinity, from which only the linear term is kept

\begin{equation}\label{Eq_B_z_expan}
 B(z)\approx B_0 \left(1+\frac{z}{L_B}\right)
 ,
\end{equation}
where $B_0=2\pi f mc/s e$ is the resonant value of the magnetic field for the frequency $f$ at the harmonic $s$, $z$ is the spatial coordinate along the line of sight with $z=0$ at $B=B_0$,  and
\begin{equation}\label{Eq_L_B}
L_B=\left(\frac{\partial B}{B\partial z}\right)^{-1}
      .
\end{equation}

Thus, the equation of GR radiation transfer through a given ($s$th) narrow gyro layer takes the form
\begin{equation}\label{Eq_Inten_GR_layer}
({\cal J}_{f,s}^\sigma)^{\mathrm{(out)}}=({\cal J}_{f,s}^\sigma)^{\mathrm{(in)}}\exp(-\tau_{s}^\sigma)+\frac{J_{f,s}^\sigma}{\tau_{s}^\sigma}\left(1-\exp(-\tau_{s}^\sigma)\right),
\end{equation}
where $({\cal J}_{f,s}^\sigma)^{\mathrm{(in)}}$ and $({\cal J}_{f,s}^\sigma)^{\mathrm{(out)}}$ are the radiation intensities entering and exiting the gyro layer, respectively; the gyro layer is characterized by integrated measures, which simplifies the theory greatly.
Eqs.\,(\ref{Eq_tau_Max_a}) and (\ref{Eq_emisZ_Max_a}) are explicitly suitable for the extension to the DDM treatment.

\newpage

\section{Radio emission from a multi-temperature plasma}
\label{S_radio_multi_T}

\subsection{Free-free emission from DEM and DDM}\label{FFfromDEM}

The e-ions component of the free-free emission depends on both DEM and DDM. The dependence on the DDM enters only via the plasma dispersion and polarization. The dependence on the DEM is more substantial.
Averaging the temperature dependent factors in Eqn.\,(\ref{Eq_emis_ff_Max_zeta}) and (\ref{Eq_abso_ff_Max_zeta}) over voxel volume, we obtain 
\begin{equation}\label{Eq_ff_comp_dem}
\left<\frac{n_e^2\ln\Lambda_C}{(k_B T)^{a}} \left(1 
+\zeta(T)\right)\right>=
 \int \frac{\xi(T)\ln\Lambda_C}{(k_B T)^{a}} \left(1
 +\zeta(T)\right) dT
            , $$$$ \quad a=1/2\ {\rm or}\ 3/2.
\end{equation}
Thus, the expressions for the e-ions emissivity and absorption coefficient take the forms
\begin{equation}\label{Eq_emis_ff_multiT_zeta}
j_{f,ff}^{\sigma}=
\frac{8e^6  n_\sigma  }{3 \sqrt{2\pi} (m c^2)^{3/2}}
 \int \frac{\xi(T)\ln\Lambda_C}{(k_B T)^{1/2}} \left(1
 +\zeta(T)\right) dT
,
\end{equation}

\begin{equation}\label{Eq_abso_ff_multiT_zeta}
\varkappa_{ff}^{\sigma}=\frac{8e^6   }{3 \sqrt{2\pi}n_\sigma  c f^2 (m )^{3/2}}
\int \frac{\xi(T)\ln\Lambda_C}{(k_B T)^{3/2}} \left(1
 +\zeta(T)\right) dT.
\end{equation}



\begin{figure*}
\resizebox{8.5cm}{!}{\includegraphics{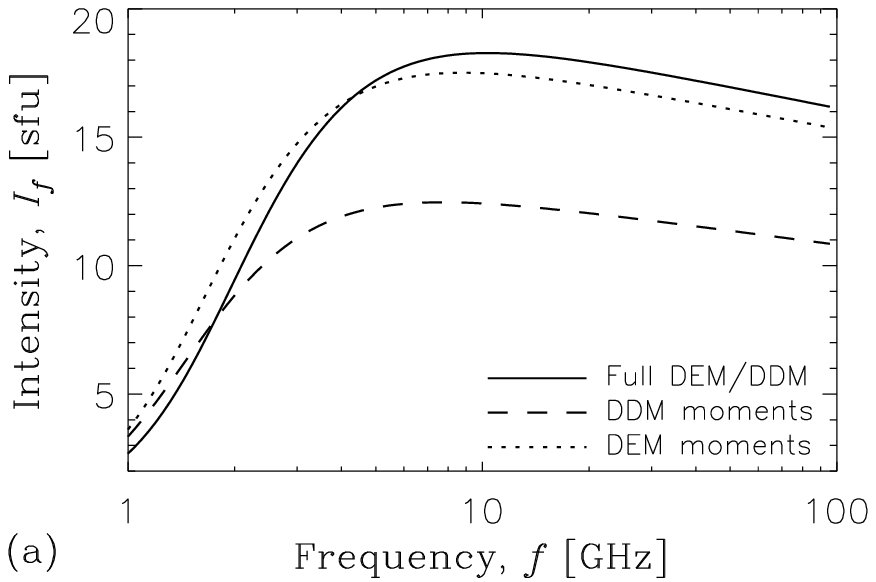}}\resizebox{8.5cm}{!}{\includegraphics{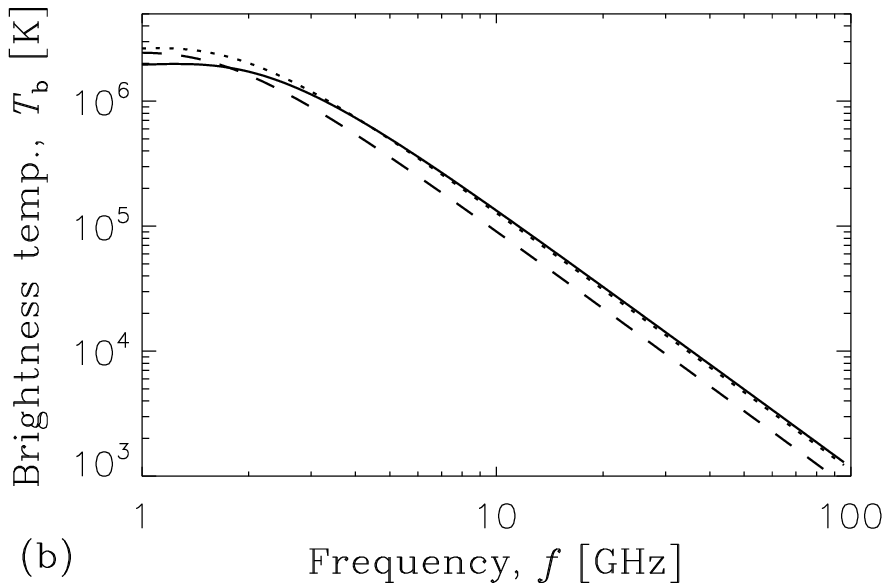}}
\caption{ Comparison of the free-free radiation spectra obtained with the full DEM/DDM treatment  (solid lines) with the  DDM-based (dashed lines) and DEM-based (dotted lines) isothermal approximations. The spectra are computed for a homogeneous hot source without magnetic field, using the new code with $\zeta$-function (coronal abundance model) and exact Coulomb logarithm. 
Simulation parameters: source area $S = 10^{20}$ $\textrm{cm}^2$, source depth $L = 1.5\times 10^{11}$ cm. The DDM and DEM distribution and the corresponding (constant) electron densities and temperatures in the isothermal models are given in Figure \protect\ref{DEMDDMexample}.}\label{FFDEMspectra}
\end{figure*}

Figure \ref{FFDEMspectra} shows an example of the free-free emission spectrum computed for the DEM and DDM distributions shown in Figure \ref{DEMDDMexample};  at the considered temperatures, plasma is fully ionized and hence the free-free emission is produced entirely due to the e-ions collisions. For comparison, the spectra computed with the  new code employing the exact Gaunt factor and the $\zeta$-function for  isothermal plasma with either DDM-based or DEM-based mean electron density and temperature are plotted. Not surprisingly, the DEM-based isothermal approximation is noticeably closer to the exact result than the DDM-based isothermal approximation. Still, the exact result employing the DEM/DDM distributions differs noticeably from that in the DEM-based isothermal approximation. In the optically thin (thick) range, the full DEM/DDM model provides higher (lower) emission intensity and brightness temperature than the isothermal models do.
Qualitatively, the effect of the DEM/DDM on the free-free emission is similar to that of the $\kappa$-distribution \citep{Fl_Kuzn_2014}; however, in the numerically defined DEM and DDM distribution we cannot find an analytical generalization of the Kirchhoff’s law unlike the $\kappa$-distribution \citep[cf. Eq. 48 in][]{Fl_Kuzn_2014}

\subsection{Gyroresonant emission from DDM}\label{GRfromDDM}
GR emissivity and optical depth depend on products $n_eT^a$, where index $a$ depends on the number of the gyroharmonics. Averaging of this products over the voxel volume is straightforward using the DDM

\begin{equation}
\label{Eq_DDM_4_GR}
 \nu_a\equiv\left<n_eT^a\right>=\frac{1}{V} \int n_e(T) T^a \frac{dV}{dT} dT = \int \nu(T)T^a dT, 
\end{equation}

Thus, the optical depth $\tau_{s}^{\sigma}$ of the gyro layer is:

\begin{equation}\label{Eq_tau_multi-T}
 \tau_{s}^{\sigma}= 
        \frac{\pi e^2 }{f mc}
       \left(\frac{k_B }{mc^2}\right)^{s-1}
       \frac{s^{2s}n_\sigma^{2s-3}\sin^{2s-2}\theta}{2^{s-1} s!(1+T_\sigma^2) } L_B \times $$$$
       [T_\sigma\cos\theta + L_\sigma\sin\theta+1]^2\times\nu_{s-1} 
             ,
\end{equation}
and, accordingly, the emissivity along the line of sight $J_{f,s}^{\sigma}$ is:
\begin{equation}\label{Eq_emisZ_multi-T}
 J_{f,s}^{\sigma}= 
        \frac{\pi e^2 f}{c}
       \left(\frac{k_B }{mc^2}\right)^{s}
       \frac{s^{2s}n_\sigma^{2s-1}\sin^{2s-2}\theta}{2^{s-1} s!(1+T_\sigma^2) } L_B \times $$$$
       [T_\sigma\cos\theta + L_\sigma\sin\theta+1]^2\times\nu_s 
             .
\end{equation}
The refraction index $n_\sigma$ and the components of the polarization vector $T_\sigma$ and $L_\sigma$ are computed for the mean plasma density defined by Eqn.\,(\ref{Eq_DM_over_ddm}).



\begin{figure}
\resizebox{8.5cm}{!}{\includegraphics{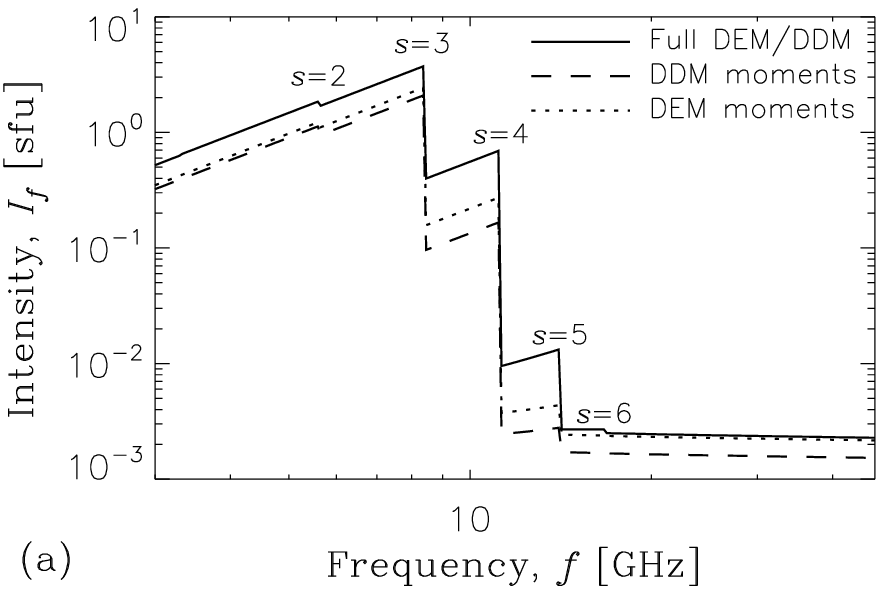}}\\
\resizebox{8.5cm}{!}{\includegraphics{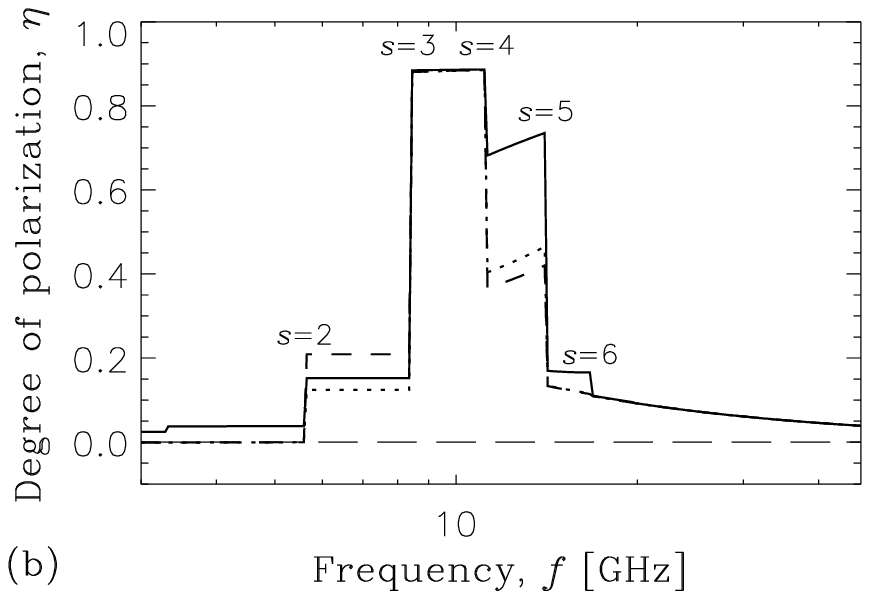}}\\
\resizebox{8.5cm}{!}{\includegraphics{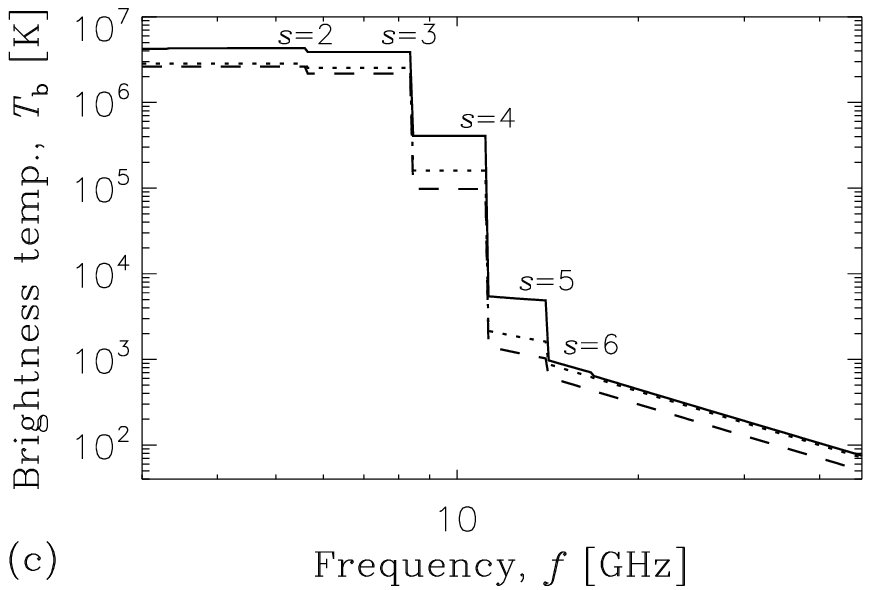}}
\caption{ Comparison of the full DDM/DEM treatment  (solid lines) and isothermal  DDM-based (dashed lines) and DEM-based (dotted lines) approximations for the gyroresonance and free-free emission. Simulation parameters: source area $S = 10^{18}$ $\textrm{cm}^2$, magnetic field decreases linearly from 1000 to 300 G over the distance of $2\times 10^9$ cm, viewing angle (constant) $\theta = 120^{\circ}$, the DDM and DEM distribution are the same in all voxels. The DDM and DEM distribution and the corresponding (constant) electron densities and temperatures in the isothermal models are presented in Figure \protect\ref{DEMDDMexample}.  The harmonic numbers $s$ printed in the panels correspond to the magnetic field strength at the farther boundary of the emission source, 1000 G.}\label{GRspectra}
\end{figure}

Figure \ref{GRspectra} shows an example of the gyroresonance emission for the DDM distribution shown in Figure \ref{DEMDDMexample}; a linear magnetic field profile is used, while all other source parameters are assumed uniform along the line of sight. The computation also accounts for the contribution of the free-free emission  using the corresponding DEM and DDM distributions. The contribution of this emission mechanism is negligible at low frequencies, while becomes significant (or even dominant) at $f\gtrsim 10$ GHz, or $s\gtrsim 6$. For comparison, the spectra computed with classical isothermal GR equations with either DEM-based or DDM-based $n$ and $T$ are shown. 
For the multithermal DDM model, the emission intensity and the brightness temperature are higher than those for the isothermal models. This is because electrons with above-average energies make a relatively larger contribution to the GR opacity. The degree of polarization can be either larger or smaller than for the isothermal plasma, because the X- and O-mode emissions at various gyro harmonics are affected by the electron temperature distribution dissimilarly. Overall, the emission properties differ measurably depending on the way the emission is computed, which justifies the use of a more advanced multithermal DEM/DDM treatment.

\section{Free-free emission from LOS DEM}

The volume-averaged definition of the DEM, Eq.\,(\ref{Eq_dem_def}), is convenient for the modeling, when the modeled DEM pertains to a given voxel. In practice, DEMs (above $\sim$0.1\,MK) are often obtained from analysis of optically thin observations, e.g., EUV and/or SXR, integrated along the line-of-sight (LOS). {It is customary to compare optically thin radio emission computed from a model employing a DEM derived from UV and/or SXR data with the observed one \citep[e.g.,][and references therein]{2001ApJ...561..396Z}. 
For example, \citet{2001ApJ...561..396Z} demonstrated that (without the $\zeta$ factor introduced in our study) the synthetic and observed radio emission are off by a noticeable factor, which they attributed to an incorrect coronal abundance of the iron. Taking the $\zeta$ factor into account would further increase this mismatch. 
}


In this section we estimate the free-free component of radio emission from the LOS DEM obtained from some (non-radio) optically thin observations. We note, that computation of the radio emission, which can be optically thin or thick depending on the frequency, is not possible from the LOS DEM without additional assumptions \citep{2019SoPh..294...23A}, because the exact solution depends on how exactly the thermal plasma is distributed along the LOS. For the same reason, computation of the GR component of radio emission from the LOS DEM is not possible either.

A reasonable estimate of the free-free emission can be obtained if we assume that the plasma is statistically uniform along the LOS such as various small portions of the LOS are characterized by the same DEM per the unit length. In this case, we can apply a uniform source solution of the transport equation with the LOS-integrated emissivity and the total optical depth obtained by LOS integration of Eqns\,(\ref{Eq_emis_ff_multiT_zeta}) and (\ref{Eq_abso_ff_multiT_zeta}):

\begin{equation}\label{Eq_emis_ff_los}
 J_{f,ff}^{\sigma}=\int\limits_{-\infty}^{\infty}j_{f,ff}^{\sigma} (z) dz= $$$$
 \frac{8e^6   }{3 \sqrt{2\pi} (m c^2)^{3/2}}
 \int \frac{\xi_l(T)\ln\Lambda_C}{(k_B T)^{1/2}} \left(1+\zeta(T)\right) dT
                    ,
\end{equation}

\begin{equation}\label{Eq_tau_ff_los}
 \tau_{ff}^{\sigma}=\int\limits_{-\infty}^{\infty}\varkappa_{ff}^{\sigma}(z) dz= $$$$
 \frac{8e^6   }{3 \sqrt{2\pi} c f^2 (m )^{3/2}}
 \int \frac{\xi_l(T)\ln\Lambda_C}{(k_B T)^{3/2}} \left(1 +\zeta(T)\right) dT
                    ,
\end{equation}
where
%

\begin{equation}
\label{Eq_los_dem_def}
 \xi_l(T)=  n_e^2(T)\frac{dz}{dT},  \quad {\rm [ cm^{-5}\, K^{-1}]}.
\end{equation}
The LOS-integrated DEM $\xi_l$ is related to the volume-averaged DEM $\xi$ introduced above as $\xi=\xi_l/L$, where $L$ is the depth of the source along the LOS. Here we adopted $n_\sigma=1$, because  neither the number density nor the magnetic field is constrained by the data (only the LOS data are available). Thus, for both X and O modes we obtain:

\begin{equation}\label{Eq_Inten_ff_DEM_LOS}
{\cal J}_{f,ff}^\sigma= \frac{J_{f,ff}^\sigma}{\tau_{ff}^\sigma}\left(1-\exp(-\tau_{ff})\right),
\end{equation}
and the total intensity (Stokes I, ${\cal J}_{f,ff}$) is equal to ${\cal J}_{f,ff}=2{\cal J}_{f,ff}^\sigma$. Eqn.\,(\ref{Eq_Inten_ff_DEM_LOS}) yields exact solution in the optically thin regime, while an estimate of the free-free emission intensity at the optically thick regime.

\begin{figure*}
\resizebox{8.5cm}{!}{\includegraphics{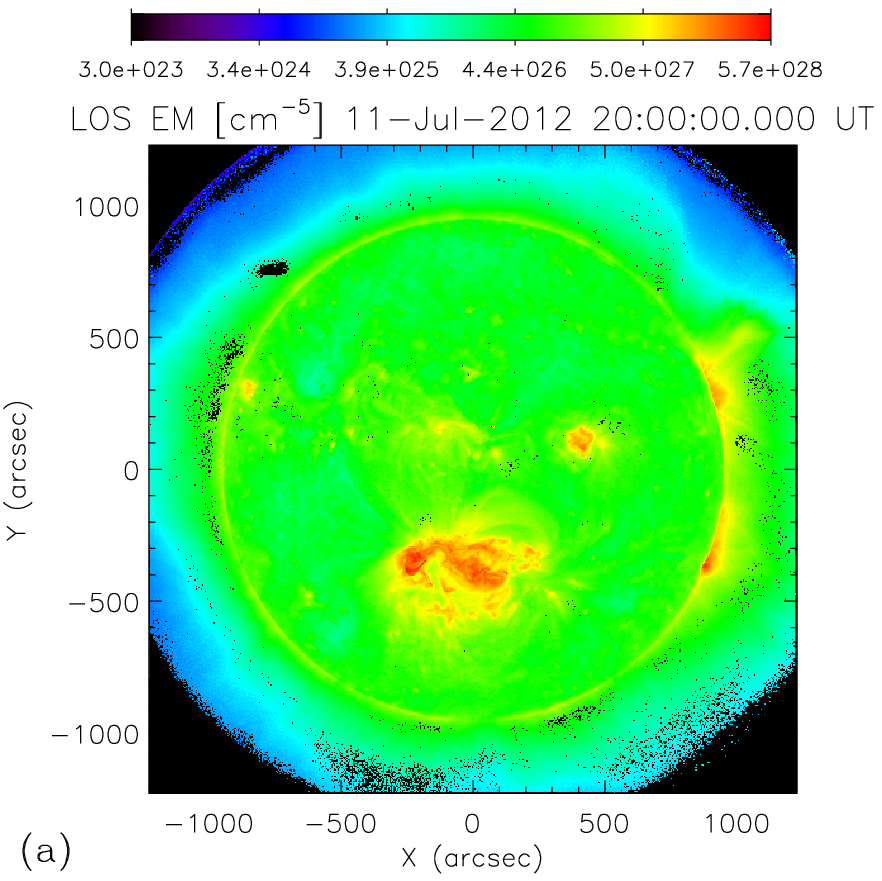}}\resizebox{8.5cm}{!}{\includegraphics{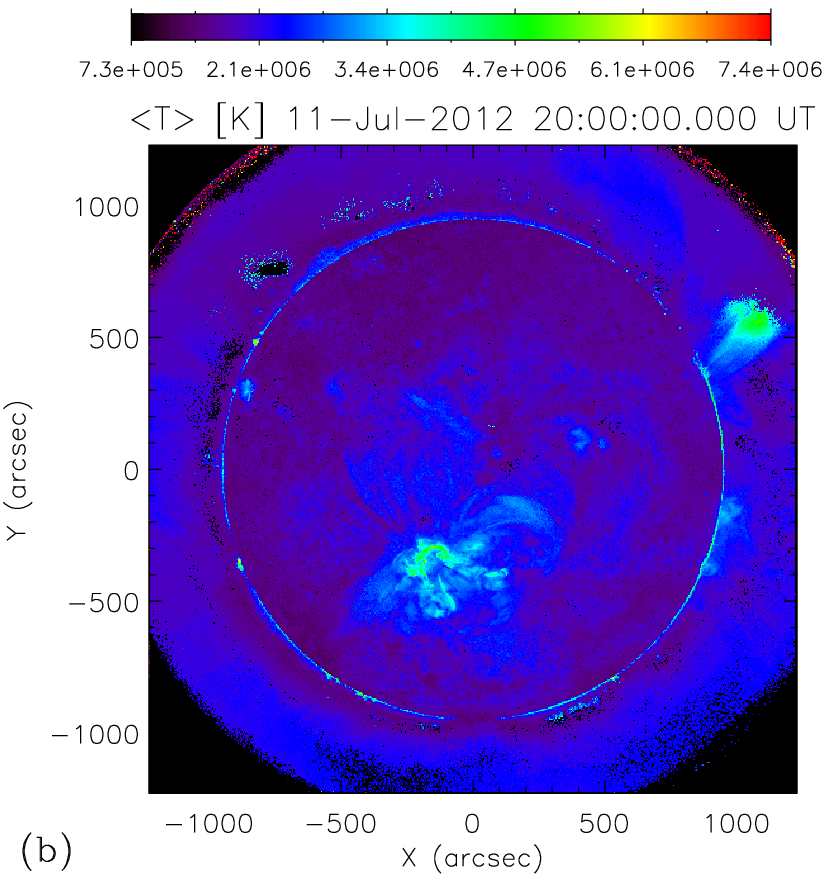}}
\caption{Maps of the total LOS-integrated emission measure (a) and average temperature (b) obtained from DEM maps available at the LMSAL Sun Today site, \url{http://suntoday.lmsal.com}. The DEM maps were inferred from the \textit{SDO}/AIA EUV observations on 11 July 2012. The color bars correspond to logarithmic (a) or linear (b) scales.}\label{EUVdata}
\end{figure*}

\begin{figure*}
\resizebox{8.5cm}{!}{\includegraphics{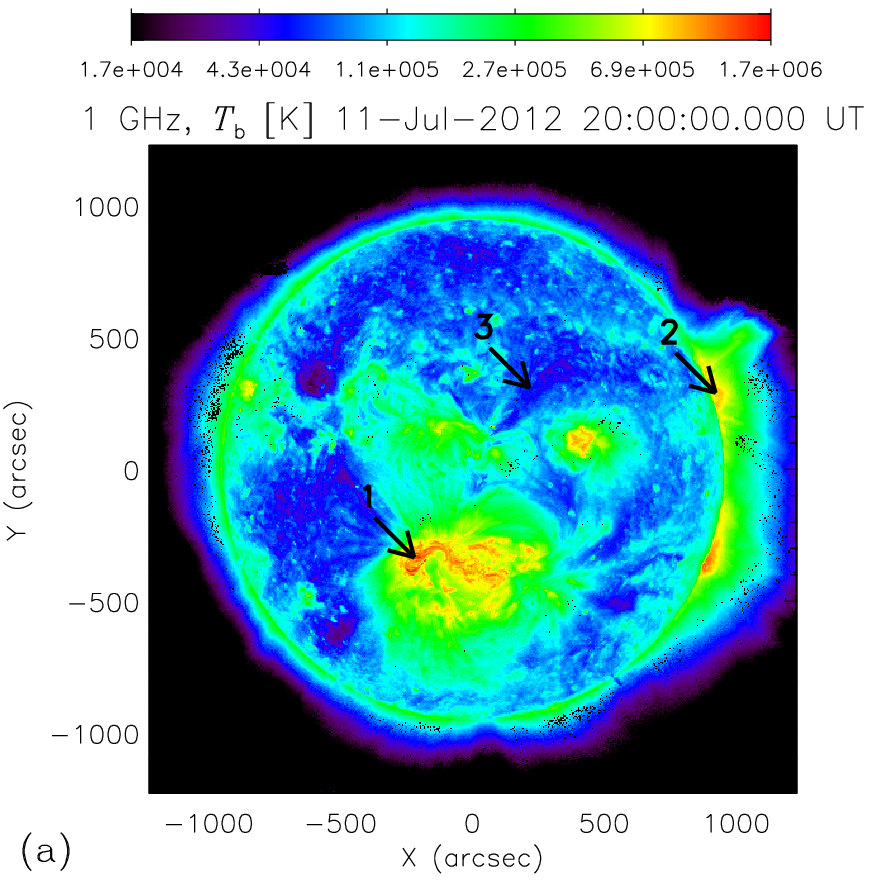}}\resizebox{8.5cm}{!}{\includegraphics{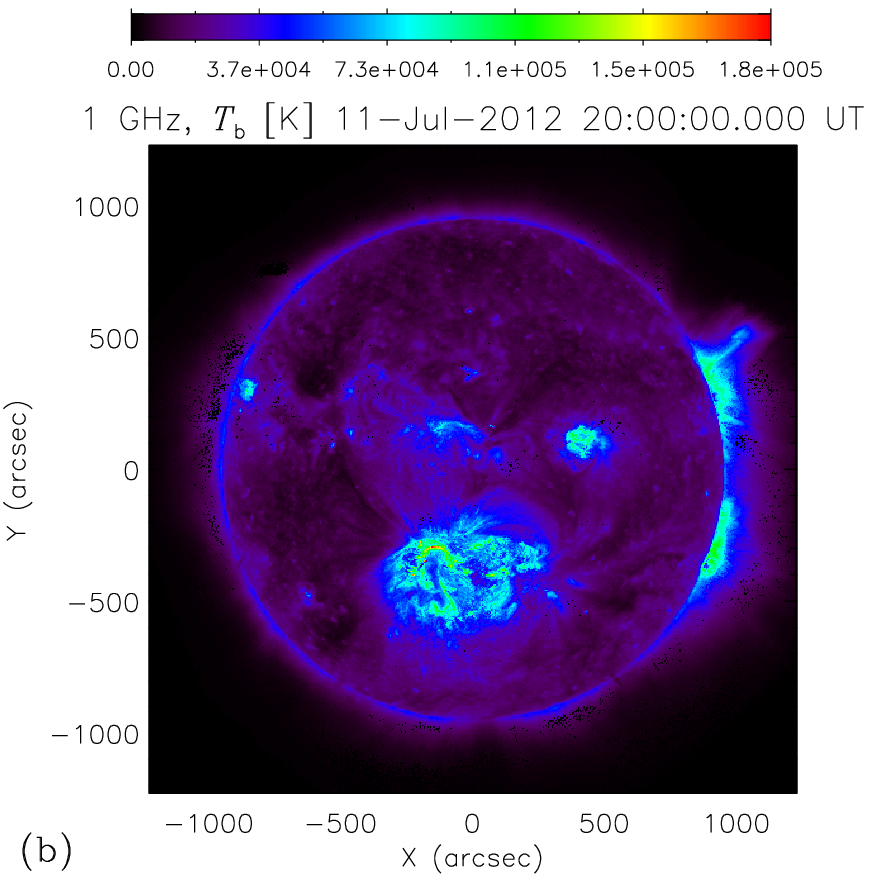}}\\
\resizebox{8.5cm}{!}{\includegraphics{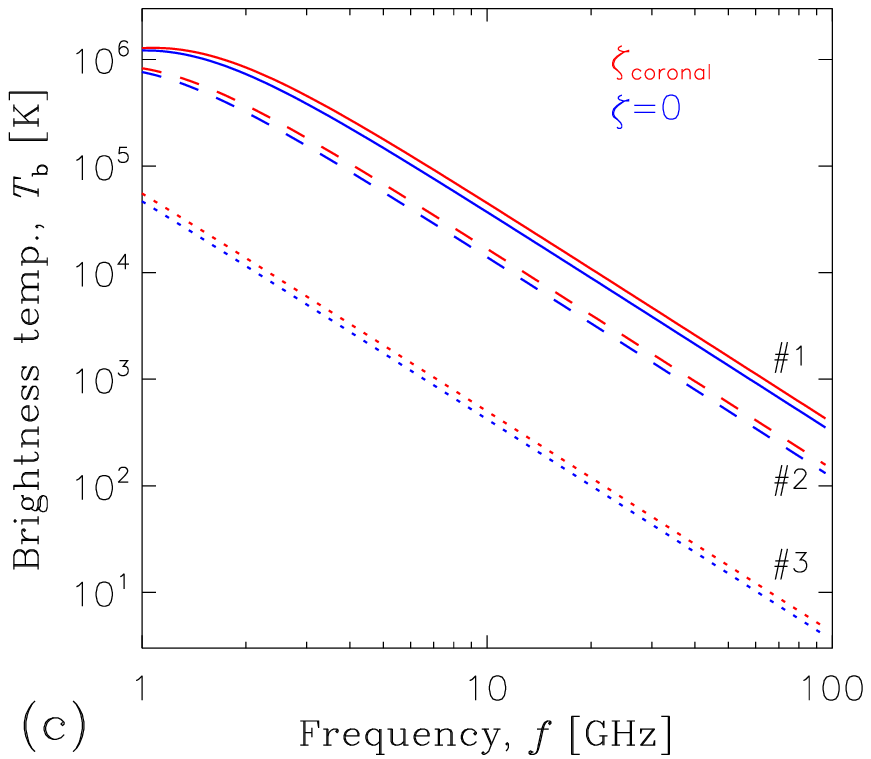}}\resizebox{8.5cm}{!}{\includegraphics{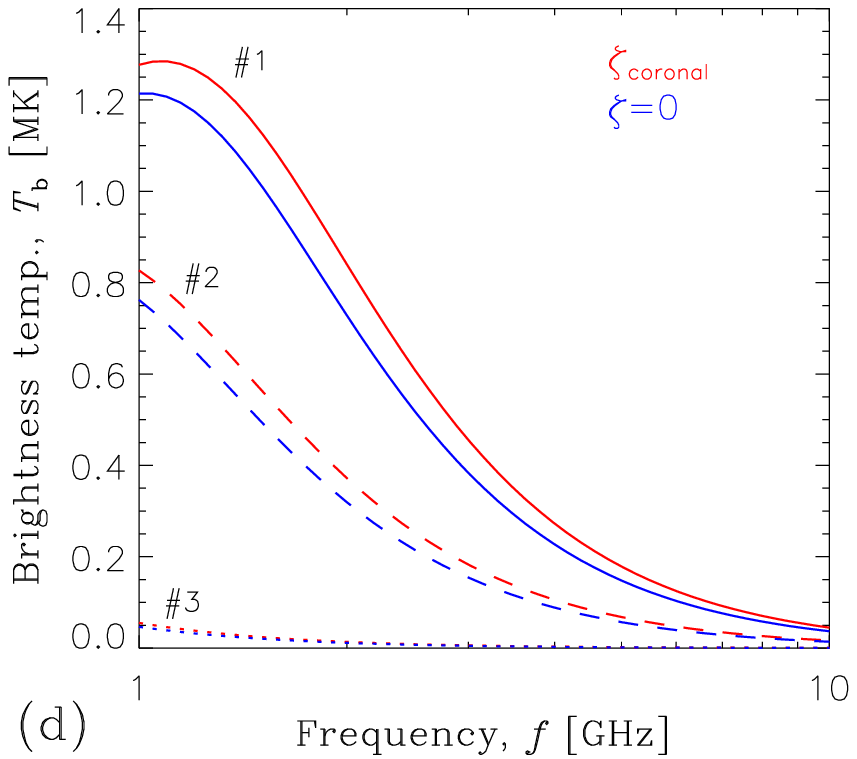}}
\caption{ (a) Synthetic map of the free-free radio emission at 1\,GHz computed from the LOS DEMs 
visualized in Figure \protect\ref{EUVdata}, {for the coronal abundance model \protect\citep{1992PhyS...46..202F}. (b) Synthetic difference radio map: emission for the coronal abundance model minus the emission for a purely hydrogen plasma ($\zeta=0$). (c-d)} Emission spectra from three representative pixels marked by arrows in panel (a), {for the coronal abundance model (red lines) and for a purely hydrogen plasma (blue lines); the selected pixels correspond to} a bright active region AR 11520 near the central meridian (\#1, solid {lines}), a bright source above the limb corresponding to the departed AR 11513 (\#2, dashed {lines}), and a quiet Sun area (\#3, dotted {lines}). {Note that panels (c) and (d) have different brightness temperature scales and frequency ranges.} The {spectra show} that the coronal free-free emission at 1\,GHz can be optically thick in active regions but optically thin in the quiet Sun. 
}\label{LOSDEMimages}
\end{figure*}

To illustrate the link between the EUV-derived DEM and the free-free emission in the radio domain, we employ data from the LMSAL ``Sun Today'' site (\url{http://suntoday.lmsal.com}). This site supplies EUV images obtained with \textit{SDO}/AIA \citep{2012SoPh..275...17L} at various passbands and 2D maps of emission measures, binned over 18 temperature ranges, derived from these EUV images using \citet{2015ApJ...807..143C} algorithm. We convert these EMs to LOS-integrated DEMs as $\xi_{l,i}=\textrm{EM}_i/\Delta T_i$, where $\Delta T_i$ is the width of $i$th temperature bin. Then we compute the corresponding volume-averaged DEMs as $\xi_i=\xi_{l,i}/L$, where  $L=10^{10}$\,cm as described above (in fact, the depth $L$ can be chosen arbitrarily as it cancels out within this single-voxel model). Finally, for each image pixel, the free-free emission is computed using the new numerical code (Section \ref{implementation}) from which both images and spectra are created. Figure \ref{EUVdata} demonstrates 
the total LOS-integrated emission measures [$\textrm{cm}^{-5}$] and average DEM-derived temperatures [K].
Figure \ref{LOSDEMimages}a shows the 2D map of the free-free radio emission at 1\,GHz {computed with the new code based on this DEM. We also computed a similar map for purely hydrogen plasma ($\zeta=0$), which looks morphologically similar to that in Figure \ref{LOSDEMimages}a and, thus, not shown. Instead, Figure \ref{LOSDEMimages}b shows a difference image of those two maps. This difference image displays rather strong mismatch between the maps; especially, where the coronal plasma is dense and hot, e.g., in active regions and the jet-like off limb feature. Panels (c,d) show spectra from three selected disk and limb locations indicated in panel (a) by numbered arrows 1--3, which are presented in both logarithmic (c) and linear (d) scales to demonstrate the spectral behavior at both large and small brightness temperatures. Note that the brightness temperature computed with the new code is larger than the old one (by up to 10\%) in both optically thin and thick regimes. This increase of the brightness temperature at the optically thick regime happens entirely due to the $\zeta$ factor in Eqns.\,(\ref{Eq_emis_ff_los}) and (\ref{Eq_tau_ff_los}.)}
We note that {in this Section} the model does not account for the gyroresonance emission. At low frequencies ($\lesssim 1$ GHz), the free-free emission from active regions can be optically thick and, thus, provide brightness temperatures of up to a few MK, i.e., similar to the observed ones, although  the gyroresonance emission can alter the shape of the emission sources strongly. The free-free emission brightness tends to decrease rapidly with frequency; hence, at $\gtrsim 3$ GHz the gyroresonance emission (not accounted here) is typically the dominant radio emission mechanism in solar active regions.

{Comparisons of the simulated and observed radio brightness are important for diagnostics of solar atmosphere; in particular, to constrain solar abundances \citep[e.g.,][]{2001ApJ...561..396Z} or to derive the LOS DEM in a broad temperature range \citep{2003ApJ...589.1054L,2008ApJ...675.1629L}. Note that corrections to the free-free emission introduced in this study may require revisiting those comparisons, because, as we have demonstrated, the free-free emission itself depends on the elemental abundances in contrast to the assumption made by \citet{2001ApJ...561..396Z}. Moreover, DEM reconstructions at low temperatures \citep{2008ApJ...675.1629L} may also require account of the electron collisions with neutral H and He atoms described in Section\,\ref{NeutralsSection}
}

\section{Code Implementation}\label{implementation}

We implemented the theory of radio emission from the multithermal multicomponent plasma described above in the computer code available {on GitHub\footnote{\url{https://github.com/kuznetsov-radio/GRFF}.}; version 1.0.0 is archived in Zenodo \citep{GRFF_2021}}. The code computes radio emission from a subset of LOSs at a range of frequencies at once, using the supplied source parameters, including the DEM and DDM distributions, specified in each voxel along each LOS. For each LOS the code numerically solves for the radiation transfer as described in Section\,\ref{S_transfer}, with the account of frequency-dependent mode coupling, which affects the polarization of emission while crossing quasi-transverse (QT) magnetic field regions \citep{1960ApJ...131..664C, 1964SvA.....7..485Z};  this is performed similarly to previous versions of our codes \citep{Fl_Kuzn_2010,Fl_Kuzn_2014}.

Inside each voxel, the way how the radiation is computed depends on what information is provided by the user. If the DEM/DDM information is provided, it is used to compute emission by default, although the user can instruct the code to use either DDM-based or DEM-based moments instead. For the voxels without the DEM/DDM information, the classical approach based on the single values of the number density and temperature is used;  in this case, the number densities of electrons, neutral hydrogen and neutral helium can be either provided explicitly by the user (e.g., following a non-LTE model), or computed by the code using the provided total number density of the 
gas within the equilibrium ionization model, as prescribed by the Saha equations\footnote{The Saha equations are used at the temperatures of $T<10^5$ K only, because at higher temperatures the plasma does not contain neutral particles.}. At low temperatures, $T<50\,000$\,K, the contribution due to electron collisions with neutrals is taken into account. 

The calling conventions, the input parameter definitions, and their format are set up such as to be seamlessly integrated into the 3D modeling and simulation tool, GX Simulator \citep{Nita_etal_2015,Nita_etal_2018}. However, the new code can be used as a stand-alone application provided that the user follows the adopted calling conventions. {The code utilizes the most recent values of the fundamental physical constants by the Bureau International des Poids and Mesures (the 2018 revision of the International System of Units)\footnote{\url{https://www.bipm.org/en/publications/si-brochure/}}.}


\section{Discussion and Conclusions}

Here we have developed a theory of radio emission in a plasma with an arbitrary set of temperatures described by a combination of the DEM and DDM distributions. This theory offers a more exact treatment of the radio emission from a multi-temperature plasma. This improvement in the accuracy of the radio emission computation due to account of the actual DEM/DDM distribution called for comparably precise account of other ingredients affecting the radio emission. Particularly, we take into account (i) elemental abundances; (ii) temperature-dependent ionization of the elements composing the plasma; (iii) exact Gaunt factor values; (iv) contribution to the emission due to collisions of the thermal electrons with neutral Hydrogen and Helium; (v) effect of the magnetic field on the free-free opacity; (vi) contribution of GR mechanism; and (vii) the frequency-dependent mode coupling in the QT layers. While refining the theory, we have identified and fixed several errors (or typos) including those in the Gaunt factor form \citep[e.g.,][]{1985ARA&A..23..169D} that propagated to other literature sources \citep[e.g.,][]{FT_2013,2020arXiv200714888N}

The theory has been implemented in a flexible computer code \citep{GRFF_2021}, which can be used to compute intensity and polarization of the radio emission from any solar or stellar atmosphere provided that the plasma remains non-relativistic. This code contains a precomputed lookup tables of the Gaunt factor and the ionization correction $\zeta$ for standard coronal and photospheric abundances in equilibrium, temperature-dependent, ionization states.

We presented several examples of radio emission computed with the new code and demonstrated that the results are generally measurably different from those obtained with the standard ``classical'' theory. The difference can be from a few percent to a factor of a few. {This finding calls for revisiting comparisons between EUV-constrained and radio-constrained DEM in the solar atmosphere \citep[e.g.,][]{2001ApJ...561..396Z,2003ApJ...589.1054L,2008ApJ...675.1629L}}.

A potential application of the new theory and the associated new code is very broad. It can be used to compute emission from a 3D model encompassing a portion of the solar/stellar atmosphere or for the entire atmosphere at a wide range of frequencies, because the code accounts for all involved physical processes at the entire range of temperatures typical for the stellar atmospheres. This new approach is needed to perform multiwavelength modeling and data analysis combining the radio data with the  EUV and SXR data, where the DEM treatment is routinely employed. Now, with the new theory and the new code, the radio emission can be computed from exactly same 3D model as used to compute the EUV/SXR emission; thus, the thermal structure of the modeled solar atmosphere can be constrained more stringently than before. Detailed modeling with the GX Simulator will be published elsewhere.

\acknowledgments
The authors are thankful to our colleagues Dale Gary, Jim Klimchuk, Maria Loukitcheva, Gelu Nita for discussions and advises.
This work was supported in part by NSF grant AST-1820613
and NASA grants
80NSSC18K0667, 
80NSSC19K0068, 
and 80NSSC18K1128 
to New Jersey Institute of Technology (GF), {by NASA grants 80NSSC18K1208 and 80NSSC20K0185 to the University of Michigan}, {and by the Ministry of Science and Higher Education of the Russian Federation}.
\newpage

\appendix
\section{Dispersion of the magnetoionic modes}
\label{S_appendix_disp}
In the cold plasma approximation, and for the frequencies far exceeding the ion plasma and cyclotron frequencies, the refraction index of an electromagnetic wave in plasma is given by \citep[e.g.,][]{1997riap.book.....Z,FT_2013}

\begin{equation}\label{N2}
n_{\sigma}^2=1-\frac{2v(1-v)}{2(1-v)-
u\sin^2\theta+\sigma\sqrt{\mathcal{D}}},
\end{equation}
where
\begin{equation}
\mathcal{D}=u^2\sin^4\theta+4u(1-v)^2\cos^2\theta,
\end{equation}
\begin{equation}
u=(f_{\mathrm{Be}}/f)^2,\qquad
v=(f_{\mathrm{pe}}/f)^2,
\end{equation}
$f$ is the wave frequency, $f_{\mathrm{pe}}=e\sqrt{n_{\mathrm{e}}/(\pi m)}$ is the electron plasma
frequency, $f_{\mathrm{Be}}=eB/(2\pi mc)$ is the electron cyclotron frequency, $n_{\mathrm{e}}$ is the number density of free electrons, and $B$ is the magnetic field strength. For X-mode, $\sigma=-1$; for O-mode, $\sigma=+1$.

The polarization vector of the wave, in the reference frame with $z$-axes along the magnetic field and the wave-vector $\mathbf{k}$ in the ($xz$)-plane, has the form
\begin{equation}
\mathbf{e}_\sigma=\frac{(T_{\sigma}\cos\theta+L_{\sigma}\sin\theta, i, -T_{\sigma}\sin\theta+L_{\sigma}\cos\theta)}
{\sqrt{1 +T_\sigma^2+L_\sigma^2}}
\end{equation}
with the parameters $T_{\sigma}$ and $L_{\sigma}$ defined as
\begin{equation}
T_{\sigma}=\frac{2\sqrt{u}(1-v)\cos\theta}{u\sin^2\theta-
\sigma\sqrt{\mathcal{D}}},
\end{equation}
\begin{equation}
L_{\sigma}=\frac{v\sqrt{u}\sin\theta+
T_{\sigma}uv\sin\theta\cos\theta}{1-u-v+uv\cos^2\theta}.
\end{equation}

Electromagnetic wave can propagate in plasma  (i.e., Equation (\ref{N2}) has a real solution corresponding to the considered wave mode) if its frequency exceeds the cutoff frequency, $f>f_{\mathrm{c}\sigma}$, where
\begin{equation}
f_{\mathrm{cO}}=f_{\mathrm{pe}},\qquad
f_{\mathrm{cX}}=\frac{f_{\mathrm{Be}}}{2}+\sqrt{f_{\mathrm{pe}}^2+\frac{f_{\mathrm{Be}}^2}{4}}.
\end{equation}

\bibliographystyle{apj}
\bibliography{AR,fleishman,Ch_Drago_2004}

\begin{thebibliography}{44}
\expandafter\ifx\csname natexlab\endcsname\relax\def\natexlab#1{#1}\fi

\bibitem[{{Alissandrakis} {et~al.}(2019){Alissandrakis}, {Bogod}, {Kaltman},
  {Patsourakos}, \& {Peterova}}]{2019SoPh..294...23A}
{Alissandrakis}, C.~E., {Bogod}, V.~M., {Kaltman}, T.~I., {Patsourakos}, S., \&
  {Peterova}, N.~G. 2019, \solphys, 294, 23

\bibitem[{{Caffau} {et~al.}(2011){Caffau}, {Ludwig}, {Steffen}, {Freytag}, \&
  {Bonifacio}}]{2011SoPh..268..255C}
{Caffau}, E., {Ludwig}, H.~G., {Steffen}, M., {Freytag}, B., \& {Bonifacio}, P.
  2011, \solphys, 268, 255

\bibitem[{{Cargill} {et~al.}(2012{\natexlab{a}}){Cargill}, {Bradshaw}, \&
  {Klimchuk}}]{2012ApJ...752..161C}
{Cargill}, P.~J., {Bradshaw}, S.~J., \& {Klimchuk}, J.~A. 2012{\natexlab{a}},
  \apj, 752, 161

\bibitem[{{Cargill} {et~al.}(2012{\natexlab{b}}){Cargill}, {Bradshaw}, \&
  {Klimchuk}}]{2012ApJ...758....5C}
---. 2012{\natexlab{b}}, \apj, 758, 5

\bibitem[{{Carlsson} {et~al.}(2016){Carlsson}, {Hansteen}, {Gudiksen},
  {Leenaarts}, \& {De Pontieu}}]{2016A&A...585A...4C}
{Carlsson}, M., {Hansteen}, V.~H., {Gudiksen}, B.~V., {Leenaarts}, J., \& {De
  Pontieu}, B. 2016, \aap, 585, A4

\bibitem[{{Cheung} {et~al.}(2015){Cheung}, {Boerner}, {Schrijver}, {Testa},
  {Chen}, {Peter}, \& {Malanushenko}}]{2015ApJ...807..143C}
{Cheung}, M. C.~M., {Boerner}, P., {Schrijver}, C.~J., {Testa}, P., {Chen}, F.,
  {Peter}, H., \& {Malanushenko}, A. 2015, \apj, 807, 143

\bibitem[{{Cohen}(1960)}]{1960ApJ...131..664C}
{Cohen}, M.~H. 1960, \apj, 131, 664

\bibitem[{{Dere} {et~al.}(2019){Dere}, {Del Zanna}, {Young}, {Landi}, \&
  {Sutherland}}]{2019ApJS..241...22D}
{Dere}, K.~P., {Del Zanna}, G., {Young}, P.~R., {Landi}, E., \& {Sutherland},
  R.~S. 2019, \apjs, 241, 22

\bibitem[{{Dere} {et~al.}(1997){Dere}, {Landi}, {Mason}, {Monsignori Fossi}, \&
  {Young}}]{1997A&AS..125..149D}
{Dere}, K.~P., {Landi}, E., {Mason}, H.~E., {Monsignori Fossi}, B.~C., \&
  {Young}, P.~R. 1997, \aaps, 125, 149

\bibitem[{{Dulk}(1985)}]{1985ARA&A..23..169D}
{Dulk}, G.~A. 1985, \araa, 23, 169

\bibitem[{{Feldman}(1992)}]{1992PhyS...46..202F}
{Feldman}, U. 1992, \physscr, 46, 202

\bibitem[{{Fleishman} {et~al.}(2002){Fleishman}, {Fu}, {Huang}, {Melnikov}, \&
  {Wang}}]{Fl_etal_2002}
{Fleishman}, G.~D., {Fu}, Q.~J., {Huang}, G.-L., {Melnikov}, V.~F., \& {Wang},
  M. 2002, \aap, 385, 671

\bibitem[{{Fleishman} \& {Kuznetsov}(2010)}]{Fl_Kuzn_2010}
{Fleishman}, G.~D., \& {Kuznetsov}, A.~A. 2010, \apj, 721, 1127

\bibitem[{{Fleishman} \& {Kuznetsov}(2014)}]{Fl_Kuzn_2014}
---. 2014, \apj, 781, 77

\bibitem[{{Fleishman} \& {Toptygin}(2013)}]{FT_2013}
{Fleishman}, G.~D., \& {Toptygin}, I.~N. 2013, Cosmic Electrodynamics.
  Astrophysics and Space Science Library; Springer NY, Vol. 388, {712 p [FT13]}

\bibitem[{{Fontenla} {et~al.}(2009){Fontenla}, {Curdt}, {Haberreiter},
  {Harder}, \& {Tian}}]{2009ApJ...707..482F}
{Fontenla}, J.~M., {Curdt}, W., {Haberreiter}, M., {Harder}, J., \& {Tian}, H.
  2009, \apj, 707, 482

\bibitem[{{Fontenla} {et~al.}(2014){Fontenla}, {Landi}, {Snow}, \&
  {Woods}}]{2014SoPh..289..515F}
{Fontenla}, J.~M., {Landi}, E., {Snow}, M., \& {Woods}, T. 2014, \solphys, 289,
  515

\bibitem[{{Gaunt}(1930)}]{1930RSPTA.229..163G}
{Gaunt}, J.~A. 1930, Philosophical Transactions of the Royal Society of London
  Series A, 229, 163

\bibitem[{{Gudiksen} {et~al.}(2011){Gudiksen}, {Carlsson}, {Hansteen}, {Hayek},
  {Leenaarts}, \& {Mart{\'{\i}}nez-Sykora}}]{2011A&A...531A.154G}
{Gudiksen}, B.~V., {Carlsson}, M., {Hansteen}, V.~H., {Hayek}, W., {Leenaarts},
  J., \& {Mart{\'{\i}}nez-Sykora}, J. 2011, \aap, 531, A154

\bibitem[{{Kakinuma} \& {Swarup}(1962)}]{1962ApJ...136..975K}
{Kakinuma}, T., \& {Swarup}, G. 1962, \apj, 136, 975

\bibitem[{{Karzas} \& {Latter}(1961)}]{1961ApJS....6..167K}
{Karzas}, W.~J., \& {Latter}, R. 1961, \apjs, 6, 167

\bibitem[{{Klimchuk} {et~al.}(2008){Klimchuk}, {Patsourakos}, \&
  {Cargill}}]{2008ApJ...682.1351K}
{Klimchuk}, J.~A., {Patsourakos}, S., \& {Cargill}, P.~J. 2008, \apj, 682, 1351

\bibitem[{{Kuznetsov} {et~al.}(2021){Kuznetsov}, {Fleishman}, \&
  {Landi}}]{GRFF_2021}
{Kuznetsov}, A., {Fleishman}, G., \& {Landi}, E. 2021, {Codes for computing the
  solar gyroresonance and free-free radio emissions}

\bibitem[{{Laming}(2015)}]{2015LRSP...12....2L}
{Laming}, J.~M. 2015, Living Reviews in Solar Physics, 12, 2

\bibitem[{{Landi} \& {Chiuderi Drago}(2003)}]{2003ApJ...589.1054L}
{Landi}, E., \& {Chiuderi Drago}, F. 2003, \apj, 589, 1054

\bibitem[{{Landi} \& {Chiuderi Drago}(2008)}]{2008ApJ...675.1629L}
---. 2008, \apj, 675, 1629

\bibitem[{{Lemen} {et~al.}(2012){Lemen}, {Title}, {Akin}, {Boerner}, {Chou},
  {Drake}, {Duncan}, {Edwards}, {Friedlaender}, {Heyman}, {Hurlburt}, {Katz},
  {Kushner}, {Levay}, {Lindgren}, {Mathur}, {McFeaters}, {Mitchell}, {Rehse},
  {Schrijver}, {Springer}, {Stern}, {Tarbell}, {Wuelser}, {Wolfson}, {Yanari},
  {Bookbinder}, {Cheimets}, {Caldwell}, {Deluca}, {Gates}, {Golub}, {Park},
  {Podgorski}, {Bush}, {Scherrer}, {Gummin}, {Smith}, {Auker}, {Jerram},
  {Pool}, {Soufli}, {Windt}, {Beardsley}, {Clapp}, {Lang}, \&
  {Waltham}}]{2012SoPh..275...17L}
{Lemen}, J.~R., {et~al.} 2012, \solphys, 275, 17

\bibitem[{{Loukitcheva} {et~al.}(2017){Loukitcheva}, {White}, {Solanki},
  {Fleishman}, \& {Carlsson}}]{2017A&A...601A..43L}
{Loukitcheva}, M., {White}, S.~M., {Solanki}, S.~K., {Fleishman}, G.~D., \&
  {Carlsson}, M. 2017, \aap, 601, A43

\bibitem[{{Nindos}(2020)}]{2020arXiv200714888N}
{Nindos}, A. 2020, arXiv e-prints, arXiv:2007.14888

\bibitem[{{Nita} {et~al.}(2015){Nita}, {Fleishman}, {Kuznetsov}, {Kontar}, \&
  {Gary}}]{Nita_etal_2015}
{Nita}, G.~M., {Fleishman}, G.~D., {Kuznetsov}, A.~A., {Kontar}, E.~P., \&
  {Gary}, D.~E. 2015, \apj, 799, 236

\bibitem[{{Nita} {et~al.}(2018){Nita}, {Viall}, {Klimchuk}, {Loukitcheva},
  {Gary}, {Kuznetsov}, \& {Fleishman}}]{Nita_etal_2018}
{Nita}, G.~M., {Viall}, N.~M., {Klimchuk}, J.~A., {Loukitcheva}, M.~A., {Gary},
  D.~E., {Kuznetsov}, A.~A., \& {Fleishman}, G.~D. 2018, \apj, 853, 66

\bibitem[{{Schmelz} {et~al.}(2012){Schmelz}, {Reames}, {von Steiger}, \&
  {Basu}}]{2012ApJ...755...33S}
{Schmelz}, J.~T., {Reames}, D.~V., {von Steiger}, R., \& {Basu}, S. 2012, \apj,
  755, 33

\bibitem[{{Scott} {et~al.}(2015){Scott}, {Grevesse}, {Asplund}, {Sauval},
  {Lind}, {Takeda}, {Collet}, {Trampedach}, \& {Hayek}}]{2015A&A...573A..25S}
{Scott}, P., {et~al.} 2015, \aap, 573, A25

\bibitem[{{Stallcop}(1974{\natexlab{a}})}]{1974ApJ...187..179S}
{Stallcop}, J.~R. 1974{\natexlab{a}}, \apj, 187, 179

\bibitem[{{Stallcop}(1974{\natexlab{b}})}]{1974A&A....30..293S}
---. 1974{\natexlab{b}}, \aap, 30, 293

\bibitem[{{van Hoof} {et~al.}(2015){van Hoof}, {Ferland}, {Williams}, {Volk},
  {Chatzikos}, {Lykins}, \& {Porter}}]{2015MNRAS.449.2112V}
{van Hoof}, P.~A.~M., {Ferland}, G.~J., {Williams}, R.~J.~R., {Volk}, K.,
  {Chatzikos}, M., {Lykins}, M., \& {Porter}, R.~L. 2015, \mnras, 449, 2112

\bibitem[{{van Hoof} {et~al.}(2014){van Hoof}, {Williams}, {Volk}, {Chatzikos},
  {Ferland}, {Lykins}, {Porter}, \& {Wang}}]{2014MNRAS.444..420V}
{van Hoof}, P.~A.~M., {Williams}, R.~J.~R., {Volk}, K., {Chatzikos}, M.,
  {Ferland}, G.~J., {Lykins}, M., {Porter}, R.~L., \& {Wang}, Y. 2014, \mnras,
  444, 420

\bibitem[{{Wedemeyer} {et~al.}(2016){Wedemeyer}, {Bastian}, {Braj{\v{s}}a},
  {Hudson}, {Fleishman}, {Loukitcheva}, {Fleck}, {Kontar}, {De Pontieu},
  {Yagoubov}, {Tiwari}, {Soler}, {Black}, {Antolin}, {Scullion}, {Gun{\'a}r},
  {Labrosse}, {Ludwig}, {Benz}, {White}, {Hauschildt}, {Doyle}, {Nakariakov},
  {Ayres}, {Heinzel}, {Karlicky}, {Van Doorsselaere}, {Gary}, {Alissandrakis},
  {Nindos}, {Solanki}, {Rouppe van der Voort}, {Shimojo}, {Kato},
  {Zaqarashvili}, {Perez}, {Selhorst}, \& {Barta}}]{2016SSRv..200....1W}
{Wedemeyer}, S., {et~al.} 2016, \ssr, 200, 1

\bibitem[{{Widing} \& {Feldman}(2001)}]{2001ApJ...555..426W}
{Widing}, K.~G., \& {Feldman}, U. 2001, \apj, 555, 426

\bibitem[{{Zhang} {et~al.}(2001){Zhang}, {Kundu}, {White}, {Dere}, \&
  {Newmark}}]{2001ApJ...561..396Z}
{Zhang}, J., {Kundu}, M.~R., {White}, S.~M., {Dere}, K.~P., \& {Newmark}, J.~S.
  2001, \apj, 561, 396

\bibitem[{{Zheleznyakov}(1962)}]{1962SvA.....6....3Z}
{Zheleznyakov}, V.~V. 1962, \sovast, 6, 3

\bibitem[{{Zheleznyakov}(1997)}]{1997riap.book.....Z}
---. 1997, {Radiation in astrophysical plasmas [in Russian]; Original Russian
  Title --- ``Izlucheniye v astrofizicheskoy plasme''}

\bibitem[{{Zheleznyakov} \& {Zlotnik}(1964)}]{1964SvA.....7..485Z}
{Zheleznyakov}, V.~V., \& {Zlotnik}, E.~Y. 1964, \sovast, 7, 485

\bibitem[{{Zlotnik}(1968)}]{1968SvA....12..245Z}
{Zlotnik}, E.~Y. 1968, \sovast, 12, 245

\end{thebibliography}

\end{document}